\documentclass[twocolumn,aps,showpacs,prb,tightenlines,amsmath,amssymb]{revtex4}
\usepackage{graphicx}
\usepackage{amssymb}
\usepackage{dcolumn}
\usepackage{amsmath}
\usepackage{bm}
\usepackage{colordvi}
\usepackage{mathrsfs}
\makeatletter

\newcommand{\Rmnum}[1]{\expandafter\@slowromancap\romannumeral #1@}
\makeatother

\begin{document}

\title{Anomalous Hall effect in semiconductor quantum wells in proximity to 
chiral $p$-wave superconductors}   

\author{F. Yang}
\affiliation{Hefei National Laboratory for Physical Sciences at
Microscale, Department of Physics, and CAS Key Laboratory of Strongly-Coupled
Quantum Matter Physics, University of Science and Technology of China, Hefei,
Anhui, 230026, China}

\author{T. Yu}
\affiliation{Hefei National Laboratory for Physical Sciences at
Microscale, Department of Physics, and CAS Key Laboratory of Strongly-Coupled
Quantum Matter Physics, University of Science and Technology of China, Hefei,
Anhui, 230026, China}

\author{M. W. Wu}
\thanks{Author to whom correspondence should be addressed}
\email{mwwu@ustc.edu.cn.}

\affiliation{Hefei National Laboratory for Physical Sciences at
Microscale, Department of Physics, and CAS Key Laboratory of Strongly-Coupled
Quantum Matter Physics, University of Science and Technology of China, Hefei,
Anhui, 230026, China}

\date{\today}

\begin{abstract} 

By using the gauge-invariant optical Bloch equation, we perform a microscopic
kinetic investigation on the anomalous Hall 
effect in chiral $p$-wave superconducting states. Specifically, the intrinsic anomalous
Hall conductivity in the absence of the magnetic field is zero as a consequence of
Galilean invariance in our description. As for the extrinsic channel, a
finite anomalous Hall current is obtained from the impurity scattering with
the optically excited normal quasiparticle current even at zero temperature. From
our kinetic description, it can be clearly seen that the excited normal
quasiparticle current is due to an induced center-of-mass momentum of Cooper
pairs through the acceleration driven by ac electric field. For the induced
anomalous Hall current, we show that the conventional skew-scattering channel in
the linear response makes the dominant contribution in
the strong impurity interaction. In this case, our kinetic description as a supplementary viewpoint
mostly confirms the results of Kubo formalism in the literature. Nevertheless, in the weak impurity
interaction, this skew-scattering channel becomes marginal and we reveal that a novel
induction channel from the Born contribution dominates the anomalous Hall
current. This novel channel, which has long been overlooked
in the literature, is due to the particle-hole asymmetry by
nonlinear optical excitation. Finally, we study the case in the chiral $p$-wave superconducting
state with a transverse conical magnetization, which breaks the Galilean
invariance. In this situation, the intrinsic anomalous Hall conductivity is no
longer zero.  Comparison of this intrinsic channel with the
extrinsic one from impurity scattering is addressed.

\end{abstract}

\maketitle 

\section{Introduction}

As one of the leading candidates for the superconducting materials with 
spontaneous time-reversal symmetry (TRS) breaking,
Sr$_2$RuO$_4$ has attracted much attention in recent
decades.\cite{Sr0,Sr1,Sr2,Sr3,Sr4,Sr5,Sr6} Considerable experiments  
including the neutron scattering\cite{NS1} and spin
susceptibility\cite{KS1,KS2,KS3} suggest that the Cooper pairing in
Sr$_2$RuO$_4$ is in the spin-triplet state. The sensitive 
superconducting transition temperature to impurities\cite{IST1,IST2} and
Josephson junction experiments\cite{Je1,Je2,Je3} are believed to reveal the odd
parity of the order parameter. Furthermore, early muon spin resonance
experiment\cite{muon} suggests that the TRS in superconducting
Sr$_2$RuO$_4$ is spontaneously broken. The most convincing indication
comes from the later observed polar Kerr effect in the absence of the
magnetic field.\cite{Kerr1}
Particularly, the Kerr angle vanishes above the superconducting transition
temperature. This clearly shows that the TRS breaking in Sr$_2$RuO$_4$ is
related to the superconductivity. These results, together with the energetic
consideration\cite{Sr5,en1,en2} and symmetry analysis,\cite{Sr2,pw1,pw2} 
reveal a chiral $p$-wave superconducting state in Sr$_2$RuO$_4$. 

Although the experimental observation of the Kerr effect in superconducting
Sr$_2$RuO$_4$\cite{Kerr1} is very convincing, the theory of the source for this
effect is not well developed. Specifically, the polar Kerr angle is proportional
to the ac anomalous Hall conductivity (AHC).\cite{qw,cm,KB1,KB2,KB3} 
Conventionally, in solids with the broken TRS, 
the intrinsic AHC arises from the anomalous velocity and hence depends on the
Berry curvature.\cite{AHC}  
Nevertheless, in superconducting Sr$_2$RuO$_4$, even though the chiral $p$-wave
character breaks TRS, for an ideal (i.e., clean single-band) model,
it is theoretically revealed that the intrinsic AHC
vanishes.\cite{qw,cm,KB1,KB2,KB3}
This result can be attributed to the consequence of the Galilean
invariance.\cite{Ga} Specifically, in systems with the translational symmetry, the
applied ac electric field only couples to the center-of-mass (CM) momentum of the
pairing electrons, independent on the relative one and hence the chiral
$p$-wave character. Therefore, mechanism with broken translational symmetry
for AHC induction is needed in chiral $p$-wave superconductor. After that, non-zero
intrinsic AHC is theoretically reported by considering the inhomogeneous effect
of the optical field.\cite{qw} Meanwhile, it is revealed that the excited collective
mode,\cite{cm} which is described as a state with oscillating
superconducting phase of the order parameter in time and
space,\cite{c1,c2,c3,c4,c5,c6,c7,c8,c9} also gives 
finite intrinsic AHC. Nevertheless, the predicted magnitudes of AHC in these two
works are much smaller than the experimentally observed one.\cite{Kerr1}

\begin{figure}[htb]
  {\includegraphics[width=5.8cm]{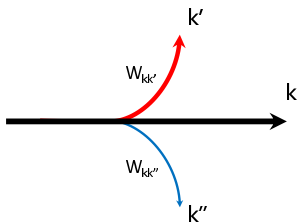}}
\caption{(Color online) Schematic of the AHC induction from impurity
  scattering. In the figure, ${\bf k'}=k{\bf e_y}=-{\bf k''}$; $W_{\bf kk'}$,
  which only depends on the relative coordinates, denotes the
  scattering probability from momentum ${\bf k}$ to ${\bf k'}$; the arrow indicates the
  quasiparticle current. This figure shows that with the optically excited normal Bogoliubov
  quasiparticle current (black arrow), the anomalous Hall current
  can be induced from impurity scattering when the scattering probability $W_{\bf
    kk'}\ne{W_{\bf kk''}}=W_{\bf k'k}$.\cite{Bo} }  
\label{figyw1}
\end{figure}

To date, the most promising theoretical explanation\cite{ep} that unveils
Kerr effect experiment in chiral $p$-wave superconductor involves extrinsic
contribution from impurity scattering.\cite{Sr5,KB1,KB2,KB3,Bo} Specifically, as  
shown in Fig.~\ref{figyw1}, with the optically excited normal Bogoliubov quasiparticle
current, the anomalous Hall current can be induced
from impurity scattering when the scattering probability $W_{\bf kk'}{\ne}W_{\bf
  k'k}$.\cite{Bo} Based on this picture, the theories of the AHC  
induction from impurity scattering in chiral $p$-wave superconductor in the
literature are developed based on either the Kubo diagrammatic
formalism\cite{KB1,KB2,KB3} or the semiclassical 
Boltzmann equation.\cite{Bo}  Specifically, in Kubo diagrammatic
treatment,\cite{KB1,KB2,KB3} with the current-current correlation in the linear
response, 
finite AHC from impurity scattering is revealed even at zero
temperature. Moreover, qualitative understanding of the AHC induction from
impurity scattering is provided. Non-zero AHC induction from the
conventional Born contribution in impurity scattering needs the
particle-hole asymmetry, which can be induced from energy
bandstructure;\cite{KB1,KB2} With the usual particle-hole symmetry, the dominant 
contribution to AHC comes from the skew  
scattering (third order of perturbation);\cite{KB1,KB2} Very recently, 
it is reported that the diffractive skew scattering\cite{KB3} (fourth order of
perturbation) also plays an important role for the AHC. As for the
semiclassical approach in the chiral $p$-wave superconductor, by 
using the scattering $T$-matrix method,  Li {\em et al.} reported the non-zero
AHC based on the Boltzmann equation of the Bogoliubov quasiparticle from an
intuitive viewpoint.\cite{Bo}  

However, there exists obvious inconsistency between
the semiclassical approach used in Ref.~\onlinecite{Bo} and the experiments for
the Kerr effect in Sr$_2$RuO$_4$.\cite{Kerr1} Specifically, the excited normal
quasiparticle current in Ref.~\onlinecite{Bo} is induced by the
drive field of the quasiparticles from the ac electric field and hence becomes negligible
around zero temperature, since the quasiparticle density decreases exponentially
with temperature. Therefore, the induced AHC from impurity scattering in their work
only dominates in the immediate vicinity of the superconducting transition
temperature and vanishes at zero temperature. This conclusion is inconsistent
with the Kerr rotation experiments where the Kerr angle achieve its maximum
around zero temperature,\cite{Kerr1} and also disagrees with the prediction from
Kubo diagrammatic formalism.\cite{KB1,KB2,KB3} The inconsistency lies in the
fact that the acceleration of the CM momentum of the Cooper pairs driven by ac
electric field is neglected in Ref.~\onlinecite{Bo} and hence no CM momentum is
excited in their work. Nevertheless, in the Kubo diagrammatic treatment, the CM
momentum, related with the vector potential ${\bf A}$, is involved in the
current-current correlation. Moreover, the induction of the CM momentum by ac
electric field is also revealed in the conventional $s$-wave superconductors from 
our microscopic kinetic description.\cite{GOBE1} It is established in the $s$-wave
superconductor that the induction of a 
large CM momentum can excite Bogoliubov quasiparticles by breaking the Cooper
pairs, i.e., forming the blocking region with the markedly suppressed anomalous
correlation,\cite{FF1,FF2,FF3,FF4,FF5,FF6,FF7,FF8,FF9} leading to the induced
quasiparticle current even at zero temperature. This quasiparticle-current
excitation through inducing the CM momentum is independent on   
the pairing symmetry\cite{FF8} and hence can also be expected in the chiral
$p$-wave superconductor. However, this acceleration effect and its further
influence on the AHC are absent in the  semiclassical approach used in Ref.~\onlinecite{Bo}. 

Furthermore, the quasiparticle correlation is overlooked in the Boltzmann
equation  used in Ref.~\onlinecite{Bo}. Nevertheless, in the $s$-wave
superconductors, the quasiparticle
correlation has been shown to play an important role in the optical response
from the approach of the Liouville equation\cite{sw1,sw2,sw3} or Bloch
equation.\cite{sw4,sw5,sw6,sw7,sw8,sw9}
Particularly, the optically excited quasiparticle correlation breaks the
particle-hole symmetry. Consequently, in the chiral $p$-wave
superconducting state, one can expect the important role of
the excited quasiparticle correlation in the AHC induction from the impurity
scattering as mentioned above. However, direct
extension of the Liouville equation\cite{sw1,sw2,sw3} or Bloch equation in
the chiral $p$-wave superconductors will lead to unphysical
conclusions. Specifically, these equations only involve the nonlinear term
${\bf A}^2$, which
leads to the pump of the quasiparticle correlation (pump effect), but the linear
effect, i.e., drive effect of particles, 
is absent. Hence, unreasonable conclusion with zero
quasiparticle current is immediately obtained. Moreover, since the pump effect
is isotropic in the momentum space, the elastic scattering is
ineffective.

Very recently, Yu and Wu pointed
out\cite{GOBE1,GOBE2} that the gauge invariance\cite{c2,c8,gi0,gi1} in the Liouville equation or Bloch equation in
the literature\cite{sw1,sw2,sw3,sw4,sw5,sw6,sw7,sw8,sw9} is
absent. Moreover, by restoring the gauge structure revealed by
Nambu,\cite{gi0,gi1} they constructed the gauge-invariant optical Bloch
equation for the $s$-wave superconducting state,\cite{GOBE1} in which both
the drive effect and the previous pump effect are included. Moreover,
the quasiparticle induction by accelerating the CM 
momentum of Cooper pairs mentioned above is also involved in their
description, and is reported to play a key role in the optical
response.\cite{GOBE1,GOBE2} Consequently, it is
natural to extend this gauge-invariant optical Bloch equation from the $s$-wave 
superconducting state to the chiral $p$-wave state, and use it to
elucidate the fundamental nature of the AHC induction from impurity scattering. 
    
However, the complex bandstructure and unclear interaction in
Sr$_2$RuO$_4$ make a detailed calculation more difficult. 
Recently, it is theoretically predicted\cite{pt1,pt2,pt3} and preliminarily
realized from the experiments\cite{pe1,pe2,pe3,pe4} that the chiral $p$-wave
superconductivity in Sr$_2$RuO$_4$ can penetrate into the normal materials
through the proximity effect. Specifically, the proximity-induced chiral $p$-wave
superconductivity is theoretically reported to 
remain long-range even in the disordered metals.\cite{pt3} After
that, direct penetration of the superconductivity from Sr$_2$RuO$_4$ into
the normal metal through a ferromagnet is observed in
Au/SrRuO$_3$/Sr$_2$RuO$_4$ junctions.\cite{pe4} Naturally, proximity
effect of chiral $p$-wave superconductivity in semiconductor quantum wells
(QWs) is expected following the success on the realization of
$s$-wave superconductivity in the semiconductor QWs via the proximity effect.\cite{sp1,sp2,sp3,sp4,sp5}
Moreover, compared to the film of metals, the
semiconductor QW possesses the simple Fermi surface and can be synthesized to be
extremely clean.\cite{QW0} Therefore, the chiral $p$-wave
superconducting QWs provide an ideal platform to study the AHC
induction. Moreover, the predictions revealed in superconducting QWs can
still shed light on the AHC induction in Sr$_2$RuO$_4$.     

In this work, with the impurity scattering included, by extending the
gauge-invariant optical Bloch equation in Ref.~\onlinecite{GOBE1} from an
$s$-wave superconducting state into the chiral $p$-wave state, we systematically
investigate the AHC induction in chiral $p$-wave superconducting QWs. In
contrast to the well established Kubo formalism,\cite{KB1,KB2,KB3} we provide a
microscopic kinetic description for the AHC induction and reveal a new AHC
induction channel from impurity scattering in our work. 
Specifically, we first demonstrate that the
intrinsic AHC without magnetic field is zero as a consequence of the Galilean
invariance in our description, in agreement with the previous
works.\cite{qw,cm,KB1,KB2,KB3} As for the
extrinsic AHC, we show that even at zero temperature, there exists optically excited
non-zero normal quasiparticle current, and a finite
anomalous Hall current from impurity scattering (Fig.~\ref{figyw1}) is obtained,
in consistence with the experiment.\cite{Kerr1} We further reveal that 
the excited quasiparticle current around zero temperature arises from the
induced CM momentum of Cooper pairs through the acceleration driven by the ac electric
field, which has been overlooked in the
previous semiclassical approach.\cite{Bo} As mentioned above, 
the induction of a large CM momentum can excite Bogoliubov quasiparticles by
breaking the Cooper pairs, i.e., forming the blocking region with markedly
suppressed anomalous correlation,\cite{FF1,FF2,FF3,FF4,FF5,FF6,FF7,FF8,FF9}
leading to the induced quasiparticle current and hence AHC from impurity scattering even
at zero temperature. 

Moreover, we find that there exist two AHC induction channels from the
impurity scattering: Channel I, conventional linear channel, where the AHC is
induced from the skew scattering in the linear response; Channel II, anomalous
nonlinear channel, where the AHC is induced from the Born contribution due to
the broken particle-hole symmetry by the nonlinearly excited quasiparticle
correlation, as mentioned above. Particularly,  
we point out that Channel I from our microscopic kinetic description, as a
supplementary viewpoint, mostly confirms the results of Kubo diagrammatic formalism of the skew
scattering. However,  we show that this conventional linear 
channel (Channel I) only dominates in the strong impurity interaction, whereas
in the weak impurity interaction, the anomalous nonlinear channel (Channel II)
makes the dominant contribution. Consequently, Channel II may
also play an important role in Sr$_2$RuO$_4$, since Sr$_2$RuO$_4$ is essentially
in the weak impurity scattering limit in order to achieve the
superconducting phase.\cite{Sr2,Sr5,IST1,IST2} Nevertheless, to the best of our
knowledge, Channel II has long been overlooked in the literature, due to the
difficulty in treating the quasiparticle 
correlation in the previous semiclassical approach\cite{Bo} or including the
nonlinear effect in Kubo diagrammatic formalism.\cite{KB1,KB2,KB3}

Furthermore, motivated by the recent observed
penetration of the superconductivity from Sr$_2$RuO$_4$ into the normal metal
through a ferromagnet in metal/ferromagnet/Sr$_2$RuO$_4$ junctions,\cite{pe4} we
also study the AHC induction in chiral $p$-wave superconducting 
QWs in the presence of a magnetization. In
our work, we consider a specific transverse conical
magnetization,\cite{TCM1,TCM2,TCM3,TCM4,TCM5} which breaks the translational
symmetry and hence the Galilean invariance. In this situation, the intrinsic AHC, attributed
from the anomalous velocity, is no longer zero.  
Comparison of this intrinsic AHC with the extrinsic one from impurity scattering
is also addressed in our work.  

Finally, we also investigate the optical response of the chiral $p$-wave
superconducting order parameter. Compared to the situation in the $s$-wave  
superconducting state,\cite{GOBE1} unique optical excitation for the
triplet-vector\cite{tv1} orientation and new feature in the collective-mode\cite{c1,c2,c3,c4,c5,c6,c7,c8,c9}
excitation are presented in our work.  

This paper is organized as follows. In Sec.~II, we introduce our model. The
analytic analysis and specific numerical results for the AHC induction are
presented in Sec.~III and~IV, respectively. We summarize in Sec.~IV.

\section{MODEL}
\label{model}

In this section, we present the Hamiltonian and set up the gauge-invariant
optical Bloch equation for the chiral $p$-wave superconducting state
by following the previous work for the $s$-wave case.\cite{GOBE1}

\subsection{Hamiltonian}

The free Bogoliubov-de Gennes (BdG) Hamiltonian of the chiral $p$-wave
superconducting QWs is given by    
\begin{equation}
\label{BdG}
H_0={\int}\frac{d{\bf r}}{2}\Psi^{\dagger}\{[\xi_{{\bf k}-e{\bf
    A}(x)\tau_3}+e\phi(x)+{\bf
      h}_{\bf r}\cdot{\bm {\tilde \sigma}}]\tau_3+{\hat \Delta}^{\bf r}_{{\bf k}}({\bf d_0})\}\Psi, 
\end{equation}
with 
\begin{equation}
\small{
{\hat \Delta}^{\bf r}_{{\bf k}}({\bf d_0})=\frac{({\bf d}_0\cdot{\bm \sigma})i\sigma_2}{2}
(\{e^{i{{\eta}}\theta_{\bf k}},e^{i\psi({\bf
      r})}\}\tau_+ -\{e^{-i{{\eta}}\theta_{\bf k}},e^{-i\psi({\bf
       r})}\}\tau_-).}
\end{equation}
Here,
$\Psi=\Big(\Psi_{\uparrow}(x),\Psi_{\downarrow}(x),\Psi^{\dagger}_{\uparrow}(x),\Psi^{\dagger}_{\downarrow}(x)\Big)^T$    
is the Nambu spinors with $x=(t,{\bf r})$ being the time-space point; $\xi_k=\varepsilon_k-\mu$ and $\varepsilon_k=\frac{k^2}{2m}$ with $m$ and $\mu$
being the effective mass and chemical potential; 
${\bf A}(x)$ and $\phi(x)$ represent the vector and scalar potential,
respectively; ${\bf d}_0=\Delta_0{\bf e_z}$ is defined as the triplet
vector;\cite{tv1} $\Delta_0$ and $\psi(x)$ stand for the magnitude and
superconducting phase of the order parameter, respectively; $\eta=\pm1$
is the chiral character; $\{~,~\}$ denotes the anticommutator;
${\bf h}_{\bf r}$ represents the magnetization with spatial dependence; ${{\bm
    {\tilde \sigma}}}={\rm diag}({\bm {\sigma}},{\bm {\sigma}^*})$; 
$\sigma_i$ and $\tau_i$ are the Pauli matrices in spin and particle-hole spaces,
respectively.

It is first revealed by Nambu that under a gauge transformation
$\Psi(x){\rightarrow}e^{i\tau_3\chi(x)}\Psi(x)$, to restore the gauge invariance of
the BdG Hamiltonian, the vector 
potential, scalar potential, and superconducting phase must transform as\cite{gi0,gi1} 
\begin{eqnarray}
eA_{\mu}&\rightarrow&eA_{\mu}-\partial_{\mu}\chi(x), \label{gaugestructure1}\\
\label{gaugestructure2}
\psi(x)&\rightarrow&\psi(x)+2\chi(x),
\end{eqnarray}
where the four vectors are $A_{\mu}=(\phi,{\bf A})$ and
$\partial_{\mu}=(\partial_t,-{\bm \nabla})$.

\subsection{Optical Bloch equation}

Following the previous work for the $s$-wave superconducting state,\cite{GOBE1}
by introducing the Wilson line\cite{Gr4,Gr5,Gr6,Gr1,Gr7} to retain the gauge  
invariance, via the nonequilibrium Green function method in the quasiparticle
approximation,\cite{Gr1,Gr2,Gr5} we set up the gauge-invariant optical Bloch equations for the
chiral $p$-wave superconducting state (refer to Appendix~\ref{AOBE}) and choose a specific gauge with
zero superconducting phase for the convenience of the physical analysis. 
The optical Bloch equations read:
\begin{eqnarray}
&&{\partial_{T}}\rho_{\bf
    k}+i\left[\left(\xi_{k}+\mu_{\rm eff}+{\bf
      h}_{\bf R}\cdot{\bm {\tilde \sigma}}+\frac{{\bf p_s}^2}{2m}\right)\tau_3+{\hat
    \Delta}_{{\bf k}}({\bf d}),\rho_{\bf
    k}\right]\nonumber\\
&&\mbox{}+\left\{\frac{{\partial_t{\bf 
 p_s}}+{\bm \nabla}_{\bf R}({\bf
      h}_{\bf R}\cdot{\bm {\tilde \sigma}})}{2}\tau_3,{\partial}_{\bf k}\rho_{\bf
    k}\right\}-\left[\frac{{\bm \nabla}_{\bf R}\cdot{\bf
    p_s}}{4m}\tau_3,\tau_3\rho_{\bf k}\right]\nonumber\\
&&\mbox{}-\left[\frac{{\bf
    p_s}}{2m}\tau_3,\tau_3{\bm \nabla}_{\bf R}\rho_{\bf
    k}\right]-\left[\frac{i}{8m}\tau_3,{\bm \nabla}^2_{\bf R}\rho_{\bf
    k}\right]=\partial_t\rho_{\bf k}|_{\rm scat},~~
\label{GOBE}
\end{eqnarray}
where  $\rho_{\bf k}$ is the density matrix in the Nambu$\otimes$spin space and
$\partial_t\rho_{\bf k}|_{\rm scat}$ represents scattering terms due to the
electron-impurity scattering; $[~,~]$ stands for the commutator.

In Eq.~(\ref{GOBE}), both the superconducting momentum ${\bf p}_s=-e{\bf
  A}+{\bm \nabla}_{\bf R}\psi$ and the effective chemical potential ${\mu}_{\rm
  eff}=e\phi+\partial_t\psi$ are gauge-invariant physical
quantities.\cite{c8,gi0,acr} They are related by the acceleration
relation:\cite{c8,gi0,acr}  
\begin{equation}
\label{acr}
\partial_t{\bf p}_s={\bm \nabla}_{\bf R}\mu_{\rm eff}+e{\bf E}.
\end{equation}

The superconducting order parameter in Eq.~(\ref{GOBE}) is written as 
\begin{equation}
\label{Higgsmode}
{\hat
    \Delta}_{{\bf k}}({\bf d})=({\bf d}\cdot{\bm \sigma})(e^{i\eta\theta_{\bf
      k}}\tau_+-e^{-i\eta\theta_{\bf k}}\tau_-)i\sigma_2,
\end{equation} 
with {\small{${\bf d}={\bf d_0}+\delta{\bf d}^{\rm HF}_{\bf
  k}$}} being the effective triplet vector and {\small{$\delta{\bf d}^{\rm HF}_{\bf k}={\rm Tr}\left[{H}^{\rm HF}_{{\bf
      k}}\tau_-i\sigma^*_2{\bm \sigma}/2\right]e^{-i\eta\theta_{\bf k}}$}}
standing for the fluctuation of the triplet vector from the
Hartree-Fock (HF) self-energy {\small{${H}^{\rm HF}_{{\bf
    k}}=\sum_{\bf k'}V_{\bf k-k'}\tau_3(\rho_{\bf k'}-\rho^0_{\bf 
  k'})\tau_3$}} due to the Coulomb interaction.\cite{GOBE1} Here, {\small{$V_{\bf
  q}={{2\pi{e^2}}}/{(q\epsilon_q)}$}} is the 2D screened Coulomb
potential; {\small{$\epsilon_q=\epsilon_0\kappa_0(1+\kappa/q)$}} with $\epsilon_0$ and
$\kappa_0$ representing the vacuum permittivity and relative dielectric
constant; the screening constant is given by
{\small{$\kappa={2me^2}/{(\epsilon_0\kappa_0)}$}} at low temperature. From
Eq.~(\ref{Higgsmode}), one can obtain the optical response of the order parameter. 

The charge-neutrality\cite{Bo,GOBE1,GOBE2,cn1,cn2,cn3} condition
requires 
\begin{eqnarray}
\label{cnc}
&&n=\sum_{\bf k}\Big\{1-\frac{\xi_{\bf k}+\mu_{\rm
    eff}}{\sqrt{(\xi_{\bf k}+\mu_{\rm eff})^2+\Delta^2_0}}+{\rm
  Tr}\left(\delta\rho^q_{\bf k}t_3\right)\Big\},~~~~~
\end{eqnarray}
with $t_3=U^{\dagger}_{\bf k}\tau_3U_{\bf k}$ and $\delta\rho^q_{\bf
  k}=\rho^q_{\bf k}-\rho^{q,0}_{\bf k}$. $\rho^q_{\bf k}=U^{\dagger}_{\bf
  k}\rho_{\bf k}U_{\bf k}$ is the density matrix in the
quasiparticle space and $n$ stands for the total electron
density. $U_{\bf k}$ represents the unitary transformation matrix from
the particle space to the quasiparticle one, which is written as 
\begin{equation}
\label{unitary}
U_{\bf k}=\left(\begin{array}{cc}
u_{\bf k}e^{i\eta\theta_{\bf k}/2}  & v_{\bf k}e^{i\eta\theta_{\bf k}/2}\sigma_1 \\
-v_{\bf k}e^{-i\eta\theta_{\bf k}/2}\sigma_1 & u_{\bf k}e^{-i\eta\theta_{\bf k}/2}
\end{array}\right),
\end{equation} 
with $u_{\bf k}=\sqrt{\frac{1}{2}+\frac{\xi_{\bf k}}{2E_{\bf k}}}$ and
$v_{\bf k}=\sqrt{\frac{1}{2}-\frac{\xi_{\bf k}}{2E_{\bf
      k}}}$. $E_k=\sqrt{\xi_k^2+\Delta^2_0}$ denotes the quasiparticle 
energy spectra. From Eq.~(\ref{cnc}), the optical response of the
effective chemical potential $\mu_{\rm eff}$ can be obtained to keep the charge
neutrality.\cite{Bo,GOBE1,GOBE2}   

The equilibrium state $\rho^0_{\bf k}$ ($\rho^{q,0}_{\bf k}$) in the particle (quasiparticle)
space is given by $\rho^0_{\bf k}=U_{\bf
  k}\rho^{q,0}_{\bf k}U^{\dagger}_{\bf k}$ with 
\begin{equation}
\label{qi}
\rho^{q,0}_{\bf
  k}=(1-\tau_3)/2+f(E_k)\tau_3.
\end{equation}
Here, $f(x)$ is the Fermi distribution. 

As for the scattering term, we mainly consider the long-range electron-impurity
scattering:\cite{Gr2}  
\begin{eqnarray}    
\label{scatinco}
&&\partial_t\rho_{\bf k}|_{\rm scat}=-\pi{n_i}\sum_{{\bf
    k'}}\sum^4_{\lambda=1}\delta(E_{k'\lambda}-E_{k\lambda})\nonumber\\
&&\mbox{}\times\left(T_{\bf
  kk'}\Gamma^{\lambda}_{\bf 
  k'}T_{\bf k'k}\Gamma^{\lambda}_{\bf k}\rho_{\bf k}-T_{\bf kk'}\rho_{\bf k'}\Gamma^{\lambda}_{\bf
  k'}T_{\bf k'k}\Gamma^{\lambda}_{\bf k}+{\rm H.c.}\right).~~~~~~~
\end{eqnarray}
Here, $n_i$ is the impurity density; $E_{{\bf
    k}\lambda}={E_{\bf k}}\delta_{\lambda,1}+{E_{\bf k}}\delta_{\lambda,2}-{E_{\bf
    k}}\delta_{\lambda,3}-{E_{\bf k}}\delta_{\lambda,4}$; 
$\Gamma^{\lambda}_{\bf k}=U_{\bf k}Q^{\lambda}U^{\dagger}_{\bf k}$ denote the
projection operators with $Q^{\lambda}={\rm
  diag}(\delta_{\lambda,1},\delta_{\lambda,2},\delta_{\lambda,3},\delta_{\lambda,4})$;
the scattering $T$-matrix $T_{\bf kk'}$ reads (refer to Appendix~\ref{Tmatrix}):
\begin{equation}
\label{TT}
T_{\bf kk'}(E)=\frac{z_iV_{\bf
  k-k'}\tau_3}{1+i\tau_3\pi\nu{z_iV_0}E/\sqrt{E^2-\Delta^2_0}},
\end{equation}
with $\nu$ and $z_i$ being the density of states and electron-impurity
interaction strength, respectively. Tuning $z_i$ can model different
scatterings including the short-range scattering. Moreover,
the prediction revealed in our work can help understanding the AHC induction
in disordered superconducting Sr$_2$RuO$_4$ samples with different impurity
potentials. Following the standard treatment of 
energy $E$ in the scattering $T$-matrix $T_{\bf kk'}(E)$ for normal 
state ($E=E_F$),\cite{Tm1,Tm2,Tm3,Tm4,Tm5,Tm6} in our work, by considering the elastic
scattering, we takes the energy $E=E_k$ in the $T$-matrix $T_{\bf kk'}(E)$,
exactly same as the previous work in superconducting state.\cite{Bo} Then,
Eq.~(\ref{TT}) becomes $T_{\bf 
  kk'}=M_{\bf kk'}e^{i\delta_k\tau_3}\tau_3$ with amplitude $M_{\bf
  kk'}=z_iV_{\bf k-k'}|\cos\delta_k|$ and phase $\delta_k=-\arctan({z_i\pi\nu{V_{0}}E_{\bf 
  k}}/{|\xi_{\bf k}|})$.

Finally, as mentioned in the introduction, with the gauge-invariant optical
Bloch equations [Eq.~(\ref{GOBE})], the drive effect of the particles from  
the drive field $\partial_t{\bf p_s}$, which excites the quasiparticle
population in the linear response, and the pump effect from $p^2_s/(2m)$ term,
which induces the quasiparticle correlation from the nonlinear effect, are both
kept. Moreover, besides the optical field, the
SG force ${\bm \nabla}_{\bf R}({\bf h}_{\bf R}\cdot{\bm {\tilde \sigma}})$ also provides a
drive field to drive the particles. 

The current during the optical response is given by:
\begin{equation}
{\bf I}=\frac{e}{2m}\int\frac{d{\bf k}}{(2\pi)^2}{\rm Tr}({\bf k}\rho_{\bf k}).
\end{equation} 

\section{ANALYTIC ANALYSIS}

In this section, from the extended optical Bloch equations [Eq.~(\ref{GOBE})], we
analytically show that the intrinsic AHC without magnetic field is zero as a
consequence of the Galilean invariance, in agreement with the 
previous works.\cite{qw,cm,KB1,KB2,KB3} As for the extrinsic channel, 
we show that the normal quasiparticle current is excited even at
zero temperature, and then, the AHC is induced from impurity scattering.  
It is further revealed that the excited normal quasiparticle current arises from
the induced CM momentum of Cooper pairs through the acceleration driven by the ac
electric field, which has been overlooked in the previous semiclassical approach.\cite{Bo}  

Moreover, two AHC induction channels from the impurity scattering are
analytically revealed: Channel I, conventional linear channel, where the AHC is
induced from the skew scattering in the linear response; Channel II, anomalous
nonlinear channel, where the AHC is induced from the Born contribution due to
the broken particle-hole symmetry by the nonlinearly excited quasiparticle
correlation.  Particularly, we point out that 
Channel I from our microscopic kinetic description mostly confirms the results
of Kubo
diagrammatic formalism of the skew scattering.\cite{KB1,KB2} Nevertheless, to
the best of our knowledge, Channel II has long been overlooked in the
literature, due to the difficulty in treating the quasiparticle
correlation in the previous semiclassical approach\cite{Bo} or including the
nonlinear effect in Kubo diagrammatic formalism.\cite{KB1,KB2,KB3} 

\subsection{Zero intrinsic AHC due to Galilean invariance}

We first analytically show that in the absence of the magnetic field,
the intrinsic AHC for the chiral $p$-wave superconducting 
state is zero as a consequence of the Galilean invariance.  
Specifically, without the magnetic field and impurity scattering included,
under a unitary transformation ${P_{\bf k}}=e^{i\eta\theta_{\bf k}/2\tau_3}$, 
Eq.~(\ref{GOBE}) is transformed into:
\begin{eqnarray}
&&{{\partial}_T\rho^s_{\bf
    k}}+i\left[\left(\begin{array}{cc} 
\xi_{\bf k}+\mu_{\rm eff}+\frac{{\bf p^2_s}}{2m} & {({\bf d}\cdot{\bm
      \sigma})}i\sigma_2\\
{({\bf d}\cdot{\bm
      \sigma})}i\sigma^*_2 &
-\xi_{\bf k}-\mu_{\rm eff}-\frac{{\bf p^2_s}}{2m}\end{array}\right),\rho^s_{\bf k}\right]\nonumber\\
&&\mbox{}+\left\{\frac{\partial_t{\bf 
 p_s}}{2}\tau_3,{\partial}_{\bf k}\rho^s_{\bf
    k}\right\}-\left[\frac{{\bm \nabla}_{\bf R}\cdot{\bf
    p_s}}{4m}\tau_3,\tau_3\rho^s_{\bf k}\right]-\left[\frac{i\tau_3}{8m},{\bm \nabla}^2_{\bf R}\rho^s_{\bf
    k}\right]\nonumber\\
&&\mbox{}-\left[\frac{{\bf
    p_s}}{2m}\tau_3,\tau_3{\bm \nabla}_{\bf R}\rho^s_{\bf
    k}\right]+\left\{\frac{\partial_t{\bf 
 p_s}}{2}\tau_3,\left[(P^{\dagger}_{\bf k}\partial_{\bf
  k}P_{\bf k}),\rho^s_{\bf k}\right]\right\}=0,\nonumber\\
\label{GA}
\end{eqnarray}
with $\rho^s_{\bf k}=P^{\dagger}_{\bf k}\rho_{\bf
  k}P_{\bf k}$ and $P^{\dagger}_{\bf k}\partial_{\bf
  k}P_{\bf k}=i\eta/(2k)\tau_3{\bf e_{\bm \theta_k}}$. 

It can be easily demonstrated that due to the fact $\{\tau_3,[\tau_3,\rho^s_{\bf
  k}]\}\equiv0$, the last term in Eq.~(\ref{GA}), which comes
from the Berry curvature, is exactly zero. Then, the optical Bloch equations
[Eq.~(\ref{GA})] for the chiral 
$p$-wave superconducting state are exactly identical as that for the $s$-wave
ones\cite{GOBE1} in the Nambu space. This can be attributed to the 
consequence of the Galilean invariance,\cite{Ga} since in systems with the
translational symmetry, the optical field only couples to the CM momentum of the
pairing electrons, independent on the relative one and hence the chiral $p$-wave
character. Consequently, it is directly concluded that the intrinsic AHC for the
chiral $p$-wave superconducting state is zero, in agreement with the previous
works.\cite{qw,cm,KB1,KB2,KB3}

\subsection{Berry curvature}
\label{Bc}

In this part, we first simplify our optical Bloch equations to demonstrate
that a finite normal quasiparticle current is excited
even at zero temperature. In the absence of the magnetic field, 
with the translational symmetry, the spatial
gradient terms in Eq.~(\ref{GOBE}) can be neglected. For the convenience of the
physical analysis, we transform the remaining optical Bloch
equations from the particle space into the quasiparticle one as:
\begin{eqnarray}
&&\partial_T\rho^q_{\bf
    k}+i[E_k\tau_3,\rho^q_{\bf
    k}]+\frac{1}{2}\{\partial_t{\bf p_s}t_3,\partial_{\bf k}\rho^q_{\bf
    k}\}\nonumber\\
&&\mbox{}+\frac{1}{2}\{\partial_t{\bf p_s}t_3,[U^{\dagger}_{\bf k}\partial_{\bf k}U_{\bf
  k},\rho^q_{\bf 
    k}]\}={{\partial}_t\rho^q_{\bf
    k}}|_{\rm scat}.
\label{Ana1}
\end{eqnarray}
Here, we have neglected the pump effect, which is isotropic in
the momentum space and hence has no direct influence on quasiparticle
current. We also neglect the HF term and effective chemical potential for simplification.  

Around zero temperature with the initial quasiparticle
distribution $f(E_k)\approx0$ in Eq.~(\ref{qi}), in the linear response, the
third term in Eq.~(\ref{Ana1}), which corresponds to the drive field to drive
the quasiparticles, is zero and hence has no contribution to quasiparticle
current. Whereas the fourth term in Eq.~(\ref{Ana1}), which arises from Berry
curvature, is finite. Consequently, from the fourth term in Eq.~(\ref{Ana1}), 
by assuming {\small{$\partial_t{\bf
  p_s}=eE_0e^{i\Omega{t}}{\bf e_x}$}} and neglecting the scattering, one has the
optically excited normal
quasiparticle current:  
\begin{equation}
{\bf I}_0=-ie^2E_0\nu{\int}d\epsilon_k\frac{(2u_kv_k)^2\varepsilon_k}{m{\Omega}E_k}{\bf e_x}, 
\end{equation}
which is non-zero even at zero temperature.

This can be understood as follows. The Berry curvature [fourth term in
Eq.~(\ref{Ana1})], overlooked in previous semiclassical approach,\cite{Bo} is
related to the acceleration of the CM momentum of Cooper pairs driven by 
the ac electric field [Eq.~(\ref{acr})], which excites finite CM momentum. 
Particularly, as mentioned in the introduction, the induction of a 
large CM momentum can excite Bogoliubov quasiparticles by breaking the Cooper
pairs, i.e., forming the blocking region with the markedly suppressed anomalous
correlation,\cite{FF1,FF2,FF3,FF4,FF5,FF6,FF7,FF8,FF9} leading to the excited
quasiparticle current even at zero temperature.

\subsection{Extrinsic AHC from impurity scattering}
\label{IS}

We next analytically show that with the optically excited normal quasiparticle current,
the anomalous Hall current can be induced from the impurity scattering. In our
study, we mainly consider the relatively weak impurity interaction $z_i\le1$. 
The impurity scattering in the quasiparticle space is given by: 
\begin{eqnarray}
\partial_t\rho^q_{\bf k}|_{\rm scat}
&=&-\pi{n_i}\nu\sum_{\xi_{k'}=\pm\xi_k}\frac{E_k}{|\xi_k|}\int\frac{d\theta_{\bf k'}}{2\pi}[W_{\bf kk'}(\rho^q_{\bf k}-\rho^q_{\bf k'})\nonumber\\
&&\mbox{}+Y_{\bf
kk'}[\tau_3,\rho^q_{\bf k'}]+{\rm H.c.}].~~\label{s1}
\end{eqnarray}
Here, $W_{\bf kk'}=t_{\bf
  kk'}{\rm Tr}(t_{\bf k'k})+Y_{\bf kk'}\tau_3$ and $Y_{\bf kk'}=t_{\bf kk'}{\rm
  Tr}(t_{\bf k'k}\tau_3)$ with $t_{\bf kk'}=U^{\dagger}_{\bf k}\tau_3U_{\bf
  k'}/2$. By
neglecting the second term in the right-hand side of Eq.~(\ref{s1}) and assuming
$\sum_{\xi_{k'}=\pm\xi_k}\approx2\delta_{k,k'}$ for simplification, 
Eq.~(\ref{s1}) approximately becomes 
\begin{equation}
\partial_t\rho^q_{\bf k}|_{\rm scat}
=-2\pi{n_i}\nu\delta_{k,k'}\frac{E_k}{|\xi_k|}\int\frac{d\theta_{\bf k'}}{2\pi}[W_{\bf
  kk'}(\rho^q_{\bf k}-\rho^q_{\bf k'})+{\rm H.c.}].\label{s2} 
\end{equation}
One can separate the
remaining scattering probability $W_{\bf kk'}=W^s_{\bf 
  kk'}+W^{a}_{\bf kk'}$ into the symmetric
part $W^s_{\bf kk'}=(W_{\bf kk'}+W_{\bf k'k})/2$, which provides the momentum
relaxation, and anti-symmetric one $W^a_{\bf kk'}=(W_{\bf kk'}-W_{\bf
  k'k})/2$, which breaks the TRS and can induce the AHC. Particularly,
the anti-symmetric scattering probability $W^a_{\bf kk'}$ at $z_i\le1$ is given by
\begin{equation}
W^a_{\bf kk'}=\sin\eta\delta\theta_{\bf kk'}|V_{\bf kk'}|^2\left(\begin{array}{cc} 
w^d_k & w^o_k\sigma_1\\
w^{o}_k\sigma_1 & w^d_k\end{array}\right),
\end{equation}
including the diagonal part $w^{d}_{k}\approx2u^2_kv^2_k\sin2\delta_k$ and off-diagonal
one $w^{o}_{k}{\approx}iu_kv_k$. Here, $\delta\theta_{\bf kk'}=\theta_{\bf k}-\theta_{\bf k'}$.

Consequently, based on the analysis in Sec.~\ref{Bc}, by assuming $\partial_t{\bf
  p_s}=eE_0(e^{i\Omega{t}}+e^{-i\Omega{t}}){\bf e_x}$ and expanding $\rho^q_{\bf
  k}=\sum_l\rho^{q,l}_{\bf k}e^{il\Omega{t}}$ in the Fourier frequency space,  
the optical Bloch equations [Eq.~(\ref{Ana1})] can be approximately simplified as:
\begin{equation}
\label{E1}
\rho^{q,1}_{{\bf k},0}=-\frac{1}{i\Omega+\tau^{-1}_{p,k}}\left[\frac{eE_0k_x(2u_kv_k)^2}{mE_k}(\rho^{q,0}_{{\bf
    k},3}+\rho^{q,2}_{{\bf
    k},3})+\Gamma^{d}_{\bf k}\right],
\end{equation}
with
\begin{eqnarray}
&&\rho^{q,2}_{{\bf k},\pm}=i\frac{eE_02u_kv_k}{2\Omega\pm{2E_k}}\left(\partial_{k_x}\rho^{q,1}_{{\bf
      k},0}\right),\label{E2}\\
&&\rho^{q,2}_{{\bf k},3}=i\frac{\Gamma^{o}_{{\bf k}-}-\Gamma^{o}_{{\bf
      k}+}}{2\Omega}, \label{E3}\\
&&\Gamma^{d}_{\bf k}=w^d_k{n_i}\nu\int\frac{2E_kd\theta_{\bf
    k'}}{|\xi_k|}\sin\eta\delta\theta_{\bf kk'}|V_{\bf kk'}|^2(\rho^{q,1}_{{\bf
    k}0}-\rho^{q,1}_{{\bf k'}0}).\nonumber\\
\\
&&\Gamma^{o}_{{\bf k}\pm}=w^o_k{n_i}\nu\int\frac{E_kd\theta_{\bf
    k'}}{|\xi_k|}\sin\eta\delta\theta_{\bf kk'}|V_{\bf kk'}|^2(\rho^{q,2}_{{\bf
    k}\pm}-\rho^{q,2}_{{\bf k'}\pm}).\nonumber\\ \label{T2}
\end{eqnarray}
Here, $\rho^{q,l}_{{\bf k},i}={\rm Tr}(\rho^{q,l}_{{\bf k}}\tau_i)$ and the
momentum relaxation rate $\tau^{-1}_{p,k}$ 
is given by {\small{$\tau^{-1}_{p,k}\approx\pi{n_i}\nu\int\frac{d\theta_{\bf
        k'}}{2\pi}\frac{E_k}{|\xi_k|}{\rm Tr}(W^s_{\bf
      kk'})(1-\cos\delta\theta_{\bf kk'})\approx4\pi{n_i}\nu|V_0|^2(1-u^2_kv^2_k)$}}.

In the right-hand side of Eq.~(\ref{E1}), the first term corresponds to the normal
quasiparticle-current excitation from the acceleration of the CM momentum
of Cooper pair, as mentioned in Sec.~\ref{Bc}. The second term denotes the
response of quasiparticle-current induction to the nonlinear optical excitation.  
The third term ($\Gamma^d_{\bf k}\propto|V_{\bf kk'}|\sin2\delta_k\propto{z^3_i}$)
stands for the skew scattering. Then, from Eq.~(\ref{E1}), the anomalous Hall current
is obtained as
\begin{equation}
I_y=\frac{2e}{m}\int\frac{d{\bf k}}{(2\pi)^2}(k_y\rho^{q,l=1}_{{\bf
    k}0})=I^{\rm I}_y+I^{\rm II}_y,
\end{equation}
with 
\begin{eqnarray}
I^{\rm I}_y&=&-\frac{2e}{m}\int\frac{d{\bf k}}{(2\pi)^2}\frac{k_y\Gamma^{d}_{\bf k}}{i\Omega+\tau^{-1}_{p,k}},\label{I1}\\
I^{\rm II}_y&=&-i\frac{2e^2}{m}E_0\int\frac{d{\bf k}}{(2\pi)^2}\frac{4u^2v^2k_xk_y}{i\Omega+\tau^{-1}_{p,k}}\frac{\Gamma^{o}_{{\bf k}-}-\Gamma^{o}_{{\bf k}+}}{2\Omega{mE_k}}.\label{II1}
\end{eqnarray}

Consequently, there exist two AHC induction channels from
the impurity scattering: Channel I, in which the anomalous Hall current
$I^{\rm I}_y$ is induced from the skew scattering ($\Gamma^d_{\bf k}\propto{z^3_i}$) in the linear
response ($\Gamma^d_{\bf k}\propto{E_0}$), and hence, we referred to this
channel as the conventional linear channel; Channel II, where the anomalous Hall current
$I^{\rm II}_y$ comes from the response of quasiparticle-current induction to the
nonlinear optical excitation [second term in the right-hand side of Eq.~(\ref{E1})],
and hence, we referred to this channel as the anomalous nonlinear channel.

For channel I, from Eq.~(\ref{I1}), one
approximately obtains
\begin{equation}
\label{IF}
I^{\rm I}_y=E_0n_i\eta\Delta_0\frac{\nu^2e^2}{2m}\int{d\left(\frac{\varepsilon_k}{\Delta_0}\right)}\frac{\Delta^4_0\varepsilon_k}{E^4_k|\xi_k|}\frac{\tau^2_{p,k}2\pi|V_{0}|^2\sin2\delta_k}{(1+i\Omega\tau_{p,k})^2}.   
\end{equation}
From above equation, it is found that $I^{\rm I}_y=0$ when the superconducting order
parameter becomes zero above the transition temperature, and $I^{\rm I}_y$ changes
sign when the chiral character $\eta\rightarrow-\eta$, in agreement with the
experiment.\cite{Kerr1} Moreover, from Eq.~(\ref{IF}),
one also finds that $I^{\rm I}_y\propto{z^3_i}$
(skew scattering), $I^{\rm I}_y\propto{E_0}$ (linear response),
$I^{\rm I}_y\propto\Delta_0$ and $I^{\rm I}_y\propto{\varepsilon_{k_F}}$. 
These dependences of Channel I from our microscopic kinetic description confirm
the results of 
the Kubo diagrammatic formalism of the skew scattering.\cite{KB1,KB2}  
Nevertheless, it is noted that the frequency dependence at high frequency in our
work ($I^{\rm I}_y\propto{\Omega^{-2}}$) is different from the Kubo diagrammatic 
formalism of the skew scattering ($I^{\rm
  Kubo}_y\propto{\Omega^{-3}}$).\cite{KB1,KB2} This frequency-dependent
difference comes from the different treatments of the
impurity scattering between our kinetic description and Kubo diagrammatic
formalism, whose detailed discussion is addressed in Sec.~\ref{summary}. 

As for channel II, from Eq.~(\ref{II1}), one finds that $I^{\rm II}_y$
is proportional to the quasiparticle
correlation ($I^{\rm II}_y\propto{\Gamma^o_{{\bf k}\pm}}$). Therefore, from Eq.~(\ref{E2}), the excited quasiparticle
correlation from the nonlinear effect is given by
\begin{equation}
\label{qco}
\rho^q_{{\bf k}\pm}=\frac{i(eE_0)^2}{i\Omega+\tau^{-1}_{p,k}}\frac{2u^3_kv^3_k}{mE_k(\Omega\pm{E_{\bf
      k}})}\left(1-6\frac{\xi_k\varepsilon_k}{E^2_k}\cos^2\theta_{\bf k}\right).
\end{equation} 
Consequently, from Eqs.~(\ref{II1}) and~(\ref{T2}), the anomalous Hall
current for Channel II is given by 
\begin{equation}
\label{Iy}
I^{\rm II}_y=i\eta\nu^2(eE_0)^3\frac{e}{m}\int\frac{d\varepsilon_k}{8m}\frac{3\Delta^6_0\varepsilon^2_k\xi_k}{|\xi_k|E^8_k\Omega(\Omega^2-E^2_k)}\frac{n_i\Gamma^{\rm
  II}_k\tau^2_{p,k}}{(1+i\Omega\tau_{p,k})^2},     
\end{equation}
with {\small{$\Gamma^{\rm II}_k=\int{d\delta\theta_{\bf kk'}}|V_{\bf kk'}|^2\sin\delta\theta_{\bf
  kk'}\sin2\delta\theta_{\bf kk'}$}}.
From above equation, one finds that  $I^{\rm II}_y=0$ when the superconducting order
parameter becomes zero above the transition temperature, and $I^{\rm II}_y$ changes
sign when the chiral character $\eta\rightarrow-\eta$, similar to Channel
I. Nevertheless, it is found that $I^{\rm II}_y\propto{E^3_0}$ (nonlinear
response) and $I^{\rm II}_y\propto{z^2_i}$ (Born contribution), differing from
Channel I. 

This anomalous nonlinear
channel from the Born contribution in the impurity scattering can be understood
as follows. During the optical response, it is noted that the excited
quasiparticle correlation from the nonlinear effect breaks the particle-hole
symmetry, since Eq.~(\ref{qco}) is not invariant under the particle-hole
transformation $\xi_{k}\rightarrow-\xi_{k}$.\cite{KB3} Moreover, it is
revealed in the previous works\cite{KB1,KB2} that in the chiral $p$-wave
superconductor, the particle-hole asymmetry from the energy bandstructure can
lead to the non-zero AHC from the Born contribution in the impurity
scattering. Similarly, the broken particle-hole symmetry through the excited
quasiparticle correlation here can also make the contribution to the AHC
induction. Nevertheless, this exotic Channel II has long
been overlooked in the literature due to the difficulty in treating the quasiparticle
correlation in the previous semiclassical approach\cite{Bo} or including the
nonlinear effect in Kubo diagrammatic formalism.\cite{KB1,KB2,KB3}

\section{NUMERICAL RESULTS}

In this section, we perform the full numerical calculation by solving
the optical Bloch equations [Eqs.~(\ref{GOBE}) and (\ref{cnc})] in the presence
of the HF and impurity scattering [$\partial_t\rho_{\bf
  k}|_{\rm scat}=U_{\bf k}\partial_t\rho^q_{\bf k}|_{\rm scat}U^{\dagger}_{\bf
  k}$ with Eq.~(\ref{s1})] terms to investigate the AHC induction from impurity
scattering in the chiral $p$-wave superconducting states. 

The specific calculation is carried out in GaAs QW in which the influence of the
spin-orbit coupling (SOC) is marginal and hence can be neglected.\cite{QW0} 
The chiral $p$-wave superconductivity in QW can be realized through the
proximity to superconducting Sr$_2$RuO$_4$.  For the optical
field, we choose a THz linear-polarized optical pulse, whose propagation
direction is assumed to be perpendicular to the QW, i.e., the ${\bf e_z}$
direction. The direction of the ac electric field is taken to be along ${\bf
  e_x}$ without loss of generality. By considering the homogeneous limit of
the optical field, from Eq.~(\ref{acr}), one has
\begin{eqnarray}
{\bf p_s}&=&(e/\Omega)E_0{\bf e_x}\sin(\Omega{t})\exp[-t^2/(2\sigma_t^2)], \label{p1}\\
\partial_t{\bf p_s}&=&eE_0{\bf
  e_x}\cos(\Omega{t})\exp[-t^2/(2\sigma_t^2)],\label{p2}
\end{eqnarray}
with $\Omega$ and $\sigma_t$ being the frequency and width of the optical
pulse. The parameters used in our computation are listed in Table~\ref{parameter}. In
our investigation, the excited normal quasiparticle current always lies in the
linear-response (i.e., small-$E_0$) regime (refer to Appendix~\ref{lrr}). 

\begin{table}[htb]
\caption{Parameters used in our calculation for GaAs QW in proximity to a chiral
  $p$-wave superconductor. $m_0$ stands for the
    free electron mass. $T$ is the temperature. $\sigma_0={e^2}/{\hbar}$.}  
\label{parameter}
  \begin{tabular}{l l l l}
    \hline
    \hline
    $m/m_0$\quad&$0.067^a$&\quad\quad$\kappa_0$\quad&$12.9^a$\\
    $n~$(cm$^{-2}$)&$2\times10^{11}$&\quad\quad$\sigma_t~$(ps)\quad&$4$\\ 
    $T~$(mK)\quad&$1$&\quad\quad$\Delta_0~$(meV)\quad&$0.206^b$\\ 
    $E_{s}~$(kV/cm)\quad&$0.1$&\quad\quad$I_0$\quad&$10^{-4}E_s\sigma_0$\\
    $q~$(nm$^{-1}$)\quad&$2\pi/60^c$&\quad\quad$\Omega/\Delta_0$\quad&$8$\\
    \hline
    \hline
\end{tabular}\\
 \quad$^a$ Ref.~\onlinecite{QW0}. \quad$^b$
  Ref.~\onlinecite{Sr1}. \quad$^c$ Ref.~\onlinecite{TCM4}.
\end{table}

We first focus on the case without the magnetic field. In this situation, with
the translational symmetry, the spatial gradient terms in Eq.~(\ref{GOBE}) can
be neglected. We show that the conventional linear channel (Channel I) only
dominates in the strong impurity interaction, while in the weak one, the
anomalous nonlinear channel (Channel II) makes the dominant contribution. 

Furthermore, 
we also study the AHC induction in the chiral $p$-wave
superconducting state with a transverse conical magnetization ${\bf h}_{\bf
  r}=h_0(\sin\theta\sin{qx},\sin\theta\cos{qx},\cos\theta)$.\cite{TCM1,TCM2,TCM3,TCM4,TCM5} Here, $\theta$ and
$q$ are the conical angle and modulation wave
vector, respectively. This transverse conical 
magnetization can be applied through the proximity
effect to the conical ferromagnet BiFeO$_3$.\cite{TCM1,TCM4} We show that with
the broken translational symmetry and hence the Galilean invariance by the
magnetization, the intrinsic AHC, attributed from the anomalous velocity, is no
longer zero. In this situation, comparison between the intrinsic AHC and extrinsic
one from impurity scattering is presented.  

Finally, the optical responses of the chiral $p$-wave superconducting
order parameter including its superconducting phase and magnitude as well as
the triplet-vector orientation are addressed.

\subsection{AHC in strong impurity interaction}

\begin{figure}[htb]
  {\includegraphics[width=8.4cm]{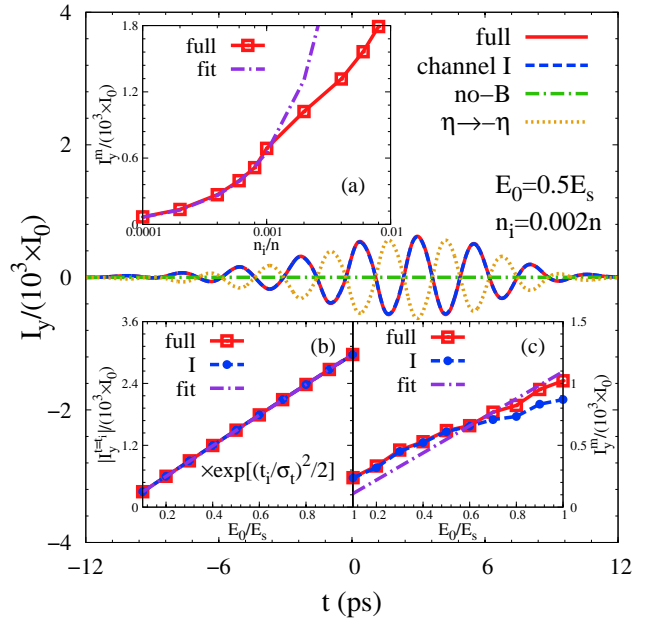}}
\caption{(Color online) Induced anomalous Hall current
  $I_y$ in the relatively strong impurity interaction ($z_i=1$). Green chain 
  line: result without the Berry curvature; Blue dashed curve: result with only
  Channel I included by
  removing $Y_{\bf kk'}$ terms and off-diagonal terms of $W^a_{\bf kk'}$ in
  Eq.~(\ref{s1}); Yellow dotted curve:
  result with the opposite chiral
  character $\eta\rightarrow-\eta$. The inset (a) shows the
  impurity-density dependence of the Hall-current maxima.
  Purple chain curve in inset (a): fitted result by using
  $I_{y}\propto{n_i}$. The inset (b) [(c)] shows the electric-field-strength
  dependence of the Hall current at $t_i=-10.13~$ps 
  (at the maxima). Purple chain curves in insets (b) and (c): fitted results by using
  $I_{y}\propto{E_0}$. } 
\label{figyw2}
\end{figure}

We first investigate the AHC induction in the relatively strong
($z_i=1$) impurity interaction in the absence of the magnetic field. 
In this situation, the temporal evolutions of the induced anomalous Hall
currents are plotted in Fig.~\ref{figyw2}. As seen from the figure, finite Hall 
currents $I_y$ with the oscillation frequency at the optical frequency $\Omega$
are observed even around zero temperature, and $I_y$ changes sign when the
chiral character $\eta\rightarrow-\eta$ (yellow dotted curve), in agreement with  
the experiment.\cite{Kerr1} The excited quasiparticle
current arises from the induced CM momentum of Cooper pairs through the
acceleration driven by the ac electric field, since by removing the Berry
curvature term in our full numerical calculation, the AHC in our work vanishes
(green chain line). This quasiparticle-current excitation by the
acceleration of the CM momentum, overlooked in the previous semiclassical
approach,\cite{Bo} confirms our analytic analysis in Sec.~\ref{Bc}.  

Additionally, in the relatively strong impurity interaction, as shown in Fig.~\ref{figyw2},
the conventional linear channel (Channel I), i.e., skew scattering, shown by blue
dashed curve which coincides with the full result (red
solid curve), makes the dominant contribution in the AHC induction. 
In this situation, at small impurity density ($n_i<0.001n$), as shown in the inset
(a) of Fig.~\ref{figyw2}, the induced anomalous Hall current (red solid curve) is linearly
enhanced by impurity density. When $n_i>0.001n$, 
the induced anomalous Hall current (red solid curve) is smaller than the fitted
result from the $n_i$-linear relation (purple chain curve). This is due to the fact
that at the large impurity density, the enhanced
momentum scattering [$\tau^{-1}_p$ in Eq.~(\ref{IF})] leads to the suppression of $I^m_y$. These
impurity-density dependences confirm our 
analytic analysis [Eq.~(\ref{IF})] and Kubo diagrammatic
formalism\cite{KB1,KB2} of the skew scattering. Particularly,  from our
analytic analysis [Eq.~(\ref{IF})], the AHC is obtained as
$\sigma^I_{xy}\approx0.308\sigma_0$ when $n_i=0.002n$, close to the value 
($\sigma^{t=t_i}_{xy}=0.3\sigma_0$) fitted from the full numerical
results at $t=-10.13~$ps [inset (b) in Fig.~\ref{figyw2}]. 
Nevertheless, $\sigma^I_{xy}$ is larger than the value
$\sigma^m_{xy}\approx0.11\sigma_0$ fitted from the full numerical result of
the Hall-current maxima [inset (c) in Fig.~\ref{figyw2}].
This difference is due to the suppressed order parameter during the optical
response, i.e., the excitation of the Higgs
mode\cite{Higg1,Higg2,Higg3,Higg4,Higg5} (refer to Sec.~\ref{ors}), which
suppresses the AHC according to our analytic analysis [Eq.~(\ref{IF})].

\begin{figure}[htb]
  {\includegraphics[width=8.4cm]{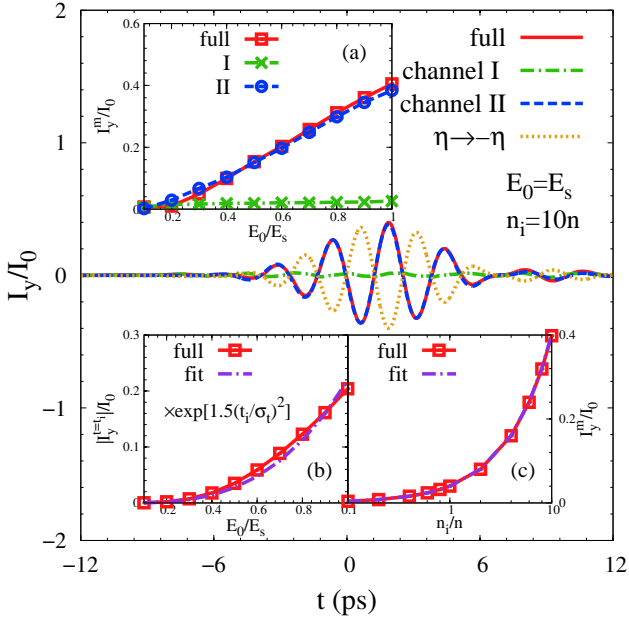}}
\caption{(Color online) Induced anomalous Hall current
  $I_y$ in the weak impurity interaction ($z_i=0.001$). Blue dashed (Green chain)
  curve: result with only Channel II (I) included by removing diagonal terms of $W^a_{\bf kk'}$
  ($Y_{\bf kk'}$ terms and off-diagonal terms of $W^a_{\bf kk'}$) in Eq.~(\ref{s1}); Yellow
  dotted curve: result 
  with the opposite chiral character $\eta\rightarrow-\eta$. The inset (a) [(b)] shows the 
  electric-field-strength dependence of the Hall current at the maxima (at
  $t_i=-4.4~$ps). The inset (c) shows the impurity-density dependence
  of the Hall-current maxima. Purple chain curve in inset (b) [(c)]: fitted result by using
  $I_{y}\propto{E^3_0}$ ($I_{y}\propto{n_i}$). } 
  
\label{figyw3}
\end{figure}

\subsection{AHC in weak impurity interaction}
\label{wi}

We next investigate the AHC induction in the weak
impurity interaction ($z_i=0.001$). The induced
anomalous Hall currents in this case are plotted in Fig.~\ref{figyw3}. As seen
from the figure, finite Hall currents with the oscillation frequency $\Omega$
are observed around zero temperature due to the acceleration of the
CM momentum of the Cooper pairs, and $I_y$ changes sign when the chiral character
$\eta\rightarrow-\eta$ (yellow dotted curve), similar to the case in the strong impurity
interaction.

Nevertheless, as shown in Fig.~\ref{figyw3} and
inset (a) of that figure, we find that Channel II (blue dashed curve)
from the Born contribution, which has long been overlooked in the literature,
makes the dominant contribution in the weak impurity interaction. Whereas
the conventional linear channel (Channel I) from the skew scattering, shown by
green chain curve, is marginal. This can be understood from the fact that the
skew scattering ($\propto{z_i^3}$) becomes marginal in the weak impurity interaction. 

From the insets (b) and (c) of Fig.~\ref{figyw3}, it is seen that the induced
Hall current $I_y\propto(eE_0)^3$ at $t_i=-4.4~$ps and the maxima
$I^{m}_y\propto{n_i}$, in agreement with our analytic analysis of Channel II
[Eq.~(\ref{Iy})]. Nevertheless, from our analytic 
analysis [Eq.~(\ref{Iy})], one has $I^{\rm 
  II}_y\approx7.5\times10^{-6}\sigma_0(E/E_s)^2$, smaller than the full numerical
results [$I_y\approx2.2\times10^{-5}\sigma_0(E/E_s)^2$] in inset (b). The
difference between the analytic and numerical results comes from the neglected
scattering term $Y_{\bf kk'}$ in Sec.~\ref{IS}, which also contributes to
Channel II. Additionally, although the Hall-current maximum $I^m_y$ [inset (a) of
Fig.~\ref{figyw3}] increases nonlinearly with the electric-field strength,
$I^m_y$ deviates from the analytic analysis [$I_y\propto(eE_0)^3$] due to the
complex nonlinear excitation and scattering process in the temporal evolution. 

\subsection{Impurity interaction strength dependence of AHC}

\begin{figure}[htb]
  {\includegraphics[width=8cm]{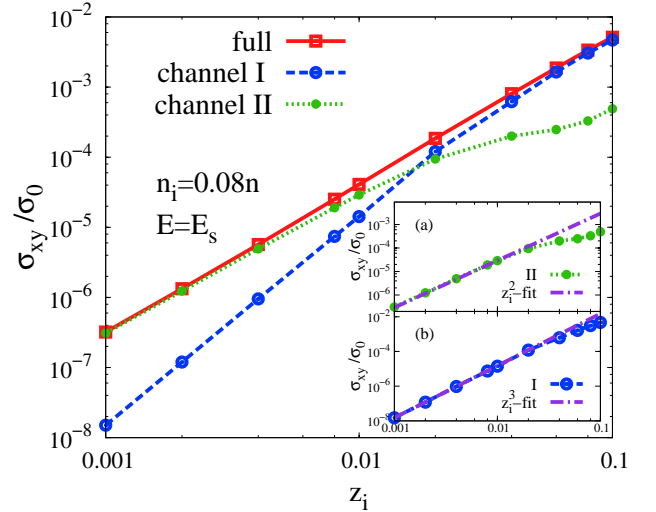}}
\caption{(Color online) AHC (fitted from the Hall-current
  maxima) versus impurity interaction strength. Green dotted (Blue dashed) curve:
  result with only Channel II (I) included by removing diagonal terms of $W^a_{\bf kk'}$
  ($Y_{\bf kk'}$ terms and off-diagonal terms of $W^a_{\bf kk'}$) in Eq.~(\ref{s1}). The inset (a) [(b)] shows the
  fitted results by using $\sigma_{xy}\propto{z^2_i}$ ($\sigma_{xy}\propto{z^3_i}$).}    
\label{figyw4}
\end{figure}

In this section, we address the impurity interaction strength dependences of the
induced AHC, which are plotted in Fig.~\ref{figyw4}. As seen
from the figure, 
Channel II (I) [green dotted curve (blue dashed curve)] makes the dominant
contribution in the weak (strong) impurity interaction when $z_i<0.02$
($z_i>0.02$). This is because that Channel I (II) arises from the skew
scattering\cite{KB1,KB2} (Born contribution in the impurity scattering),  
and hence, based on the Kubo formalism,\cite{KB1,KB2} the induced AHC is
proportional to $z^3_i$ ($z^2_i$), in agreement with our numerical result shown
in the inset (a) [(b)] of Fig.~\ref{figyw4} and analytic one in Eq.~(\ref{IF})
[Eq.~(\ref{Iy})]. It is noted that in the insets (a) and (b), the smaller
numerical results than the analytic ones when $z_i>0.04$ is due to the
suppression from the enhanced momentum relaxation.

\subsection{Intrinsic channel by introducing transverse conical magnetization}

Finally, we show that in the presence of the transverse conical
magnetization, the intrinsic AHC is no 
longer zero. Specifically, with the transverse conical
magnetization, the translational symmetry is
broken. In this situation, considering the fact that the magnetization varies
smoothly with the length scale $l=2\pi/q\gg\xi$ ($\xi$ denotes the superconducting
coherence length), we treat the spatial derivative of
${\bf h_R}$ perturbatively. This treatment can be
simplified\cite{SOC1,SOC2,SOC3} by transforming the optical Bloch 
equations [Eq.~(\ref{GOBE})] into the helical space (in which the spin axis
is parallel to ${\bf h_R}$) through a unitary transformation matrix $Q_{\bf
  R}=e^{-{i({\bm \alpha_{\bf R}}{\cdot}{\bm {\tilde \sigma}})\tau_3}}$ via keeping non-Abelian
gauge invariance\cite{Gr6} (more details
of the derivation refer to Appendix~\ref{AOBE}). Here, ${\alpha_{\bf 
    R}}=\frac{\pi}{2}(\sin(\theta/2)\sin(qx),\sin(\theta/2)\cos(qx),\cos(\theta/2))$. 
In the helical space, the magnetization becomes homogeneous. In this
    situation, following the previous works,\cite{SOC1,SOC2,SOC3}
one can neglect the spatial derivatives and the
optical Bloch equations read: 
\begin{eqnarray}
&&{{\partial}_T\rho^h_{\bf
    k}}+i\left[\left(\begin{array}{cc}  
\xi_k+\mu_{\rm eff}+h_0{{\sigma}}_z & ({\bf d}^h\cdot{\bm \sigma})e^{i{{\eta}}\theta_{\bf k}}\\
({\bf d}^h\cdot{\bm \sigma})e^{-i{{\eta}}\theta_{\bf k}} &
-\xi_k-\mu_{\rm eff}-h_0{\sigma}_z
\end{array}\right),\rho^h_{\bf k}\right]\nonumber\\
&&\mbox{}+\frac{1}{2}\left\{\partial_t{\bf 
 p_s}\tau_3,{{\partial}_{\bf k}\rho^h_{\bf
    k}}\right\}+\frac{1}{2}\left\{[({\bf
      A_s}\cdot{\bm {\tilde \sigma}}){\bf e_x},h_0{\tilde
      \sigma}_z],{{\partial}_{\bf k}\rho^h_{\bf
    k}}\right\}\nonumber\\
&&\mbox{}+i\left[\frac{{\bf
        p_s}^2}{2m}\tau_3,\rho_{\bf k}\right]-\frac{i}{2}\left\{\frac{k_x}{m}\tau_3,\left[({\bf
      A_s}\cdot{\bm {\tilde \sigma}})\tau_3,\rho^h_{\bf k}\right]\right\}\nonumber\\
&&\mbox{}+\frac{i}{2}\left[\frac{\bf
  p_s\cdot{e_x}}{m}\tau_3,\left\{({\bf
      A_s}\cdot{\bm {\tilde \sigma}}),\rho^h_{\bf k}\right\}\right]={{\partial}_t\rho_{\bf
    k}}|_{\rm scat},
\label{Em2}
\end{eqnarray}
where ${\bf A_s}=\partial_x{\bm \alpha}_{\bf
  R}=\frac{q\pi}{2}\sin(\frac{\theta}{2})(\cos(qx),-\sin(qx),0)$;
$\rho^{h}_{\bf k}=Q^{\dagger}_{\bf R}\rho_{\bf k}Q_{\bf R}$ and $({\bf
  d}^h\cdot{\bm \sigma})=e^{i({\bm \alpha_{\bf R}}{\cdot}{\bm {\sigma}})}({\bf
  d}\cdot{\bm \sigma})e^{-i({\bm \alpha_{\bf R}}{\cdot}{\bm {\sigma}})}$ are
the density matrix and effective triplet vector in the helical space, respectively.

In Eq.~(\ref{Em2}), the fourth term corresponds to the drive field from the SG
force. The sixth term acts as an effective SOC from the spin-rotational
transformation.\cite{SOC1,SOC2,SOC3,Gr6}  The seventh term, which is the
coupling between the optical field and magnetization, acts as an excited 
effective in-plane 
magnetic field along ${\bf A}_s$. In this situation, with ${\bf p_s}$ and $\partial_t{\bf
  p_s}$ given in Eqs.~(\ref{p1}) and (\ref{p2}), we perform the numerical
calculation by solving the optical Bloch equations [Eqs.~(\ref{Em2}) and
(\ref{cnc})] in the presence of the impurity scattering. We find that the
induced effective chemical potential and the induced current have no spatial
dependence (not shown). This indicates the homogeneous condition (${\bm \nabla}_{\bf R}\mu_{\rm eff}=0$)
to use Eqs.~(\ref{p1}) and (\ref{p2}) is still satisfied here. 

\begin{figure}[htb]
  {\includegraphics[width=7.8cm]{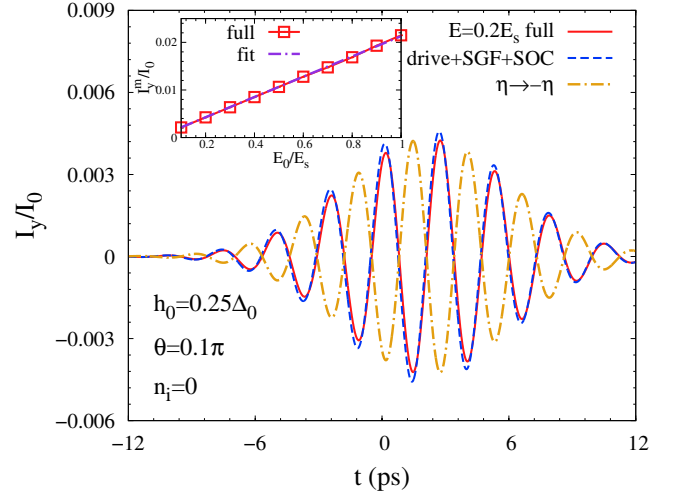}}
\caption{(Color online) Induced anomalous Hall current $I_y$ in the presence of
  the transverse conical magnetization in the clean limit. Blue dashed curve:
  result with only the drive effect, SOC and SG force included by setting ${\bf
    p}_s=0$. Yellow chain curve: result with the opposite
  chiral character $\eta\rightarrow-\eta$. The inset shows electric-field-strength
  dependence of the Hall-current maxima. Purple chain curve in the inset:
  fitted result by using $I_y\propto{E_0}$. }    
\label{figyw5}
\end{figure}

We first focus on the clean limit. In this situation, finite anomalous Hall
currents with the oscillation frequency at the optical frequency $\Omega$, 
plotted in Fig.~\ref{figyw5}, are observed. Particularly, the anomalous Hall
current comes from the linear response, since $I^m_y$ is linearly proportional
to $E_0$, as shown in the inset of Fig.~\ref{figyw5}.  

The induced anomalous Hall current here is attributed to the broken
translational symmetry and hence the Galilean invariance by the transverse conical
magnetization. This can also be understood as follows.
Through the SOC, i.e., momentum-dependent effective magnetic
field, the spin-polarized quasiparticles are induced with a net
quasiparticle momentum driven by the ac electric field.
Consequently, the SG force drives the spin-polarized quasiparticles to form the
net anomalous current via the anomalous velocity (i.e., Berry
curvature). This analysis is demonstrated in Fig.~\ref{figyw5}, since the result
with only the drive effect, SG force and SOC included (blue dashed curve) almost
coincides with the full one (red solid curve).   

Moreover, based on the analysis above, we give the analytical formula for the
anomalous Hall current induced by the transverse conical
magnetization in the linear response (refer to Appendix~\ref{Din}), which is
approximately given by: 
\begin{equation}
\label{TCMin}
I^t_y=16E_0\sigma_0\pi^3\eta{\sin^4\left(\frac{\theta}{2}\right)}\frac{h^4_0\epsilon^2_q\Delta^2_0}{\Omega^8}\int{d\left(\frac{\epsilon_{k}}{\Delta_0}\right)}\left(\frac{\Delta_0}{E_k}\right)^3.     
\end{equation}
From above equation, it is found that $I^t_y$ changes sign when the
chiral character $\eta\rightarrow-\eta$, in consistence with the full numerical
result (brown chain curve) in Fig.~\ref{figyw5}, and $I^t_y=0$ when the
superconducting order parameter becomes zero above the transition
temperature. This indicates that the induced intrinsic anomalous Hall effect is
exactly related to the chiral $p$-wave superconductivity, similar to the
extrinsic AHC induction from impurity scattering.  Furthermore, by restoring
the translational symmetry via setting conical angle $\theta=0$ or modulation
wave vector $q=0$, one has $I^t_y=0$ from Eq.~(\ref{TCMin}), in consistence with
the Galilean invariance. Additionally, from Eq.~(\ref{TCMin}), the AHC is obtained as
$\sigma^{t}_{xy}\approx2.02\times10^{-6}\sigma_0$, close
to the value ($\sigma^{\rm fit}_{xy}=2.1\times10^{-6}\sigma_0$) fitted
from the full numerical results (inset in Fig.~\ref{figyw5}). 

\subsection{Comparison between extrinsic and intrinsic channels}

\begin{figure}[htb]
  {\includegraphics[width=8.2cm]{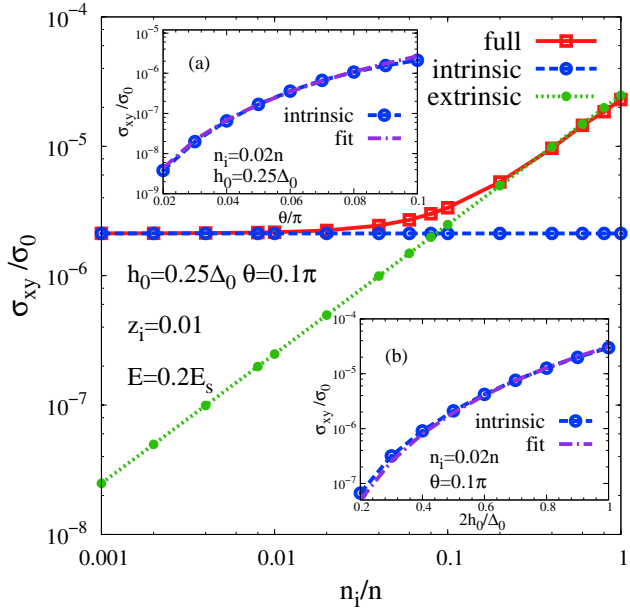}}
\caption{(Color online)  AHC (fitted from the Hall-current
  maxima) versus impurity density in the presence of the transverse conical
  magnetization and impurity 
  scattering. Blue dashed (Green dotted) curve: result with only the intrinsic
  (extrinsic) AHC included by removing $W^a_{\bf kk'}$
  and $Y_{\bf kk'}$ terms in Eq.~(\ref{s1}) [SG force in Eq.~(\ref{Em2})]. The inset
  (a) [(b)] shows conical angle $\theta$ (magnetization strength $h_0$)
  dependence of the intrinsic AHC. Purple chain curve in 
  the inset (a) [(b)]: fitted result by using $I_y\propto\sin^4(\theta/2)$ ($I_y\propto{h^4_0}$).}   
\label{figyw6}
\end{figure}

In this part, we compare the extrinsic AHC induction channel
from the impurity scattering with the intrinsic one from transverse conical
magnetization. As shown in Fig.~\ref{figyw6},
the intrinsic channel from magnetization (blue dashed curve), which is nearly
invariant with the increase of the impurity density, dominates at small impurity
density. The extrinsic induction channel from the impurity scattering (green
dotted curve), is enhanced by the increase of impurity density and makes the
dominant contribution at large impurity density. 

Moreover, the intrinsic channel from the transverse conical magnetization can
be enhanced though the enhancement of the 
translational symmetry breaking by increasing the strength $h_0$ and conical
angle $\theta$, as shown in the insets (a) and (b) of
Fig.~\ref{figyw6}, respectively, in good agreement with the analytic analysis
[Eq.~(\ref{TCMin})]. This provides a scheme to experimentally distinguish these
two channels. 

\subsection{Optical response of the superconducting order parameter}
\label{ors}

\begin{figure}[htb]
  {\includegraphics[width=8.0cm]{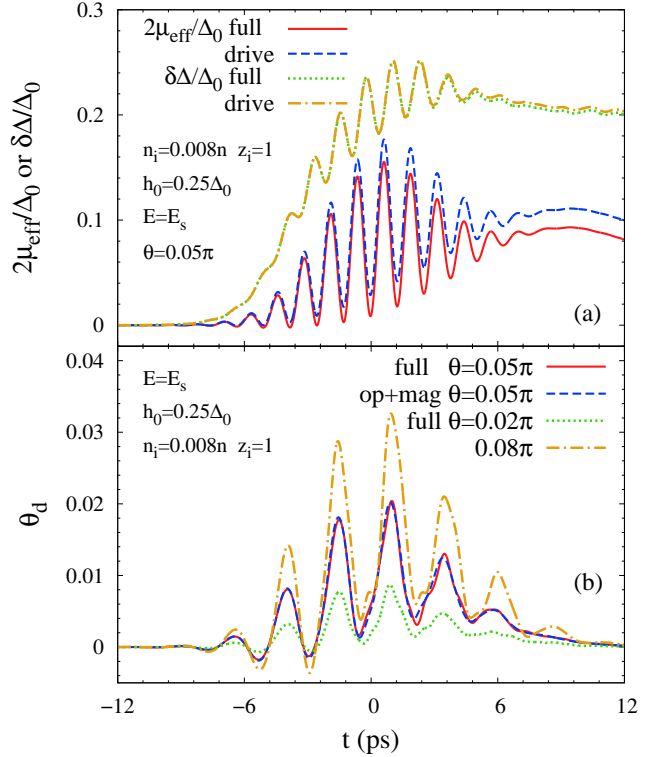}}
  \caption{(Color online) Temporal evolutions of the effective chemical
    potential, magnitude of the triplet vector $\delta\Delta=\sum_{\bf k}(|{\bf d_0}|-|{\bf
      d}_{\bf k}|)/(\sum_{\bf k}1)$ in (a) and longitude angle of the triplet vector 
    $\theta_{\bf d}=\sum_{\bf k}\theta_{\bf d_k}/(\sum_{\bf k}1)$ in (b). Blue dashed and
    yellow chain curves in (a): results
    with only the drive effect included. Blue
    dashed curve in (b): result without the SOC and SG force in Eq.~(\ref{Em2}). }   

\label{figyw7}
\end{figure}

Finally, we study the optical response of the superconducting order
parameter. We first investigate the effective
chemical potential $\mu_{\rm eff}$ in Fig.~\ref{figyw7}(a), which is related to
the phase of the superconducting order parameter (known as 
the collective mode\cite{c1,c2,c3,c4,c5,c6,c7,c8,c9,GOBE1}). Compared to the
situation for the $s$-wave superconducting state in the previous
work\cite{GOBE1} where the pump and drive effects both play important roles, for
the chiral $p$-wave case, the drive effect (blue dashed curve) dominates the
excited oscillation of $\mu_{\rm eff}$ with frequency $2\Omega$ (red solid curve).  
This difference lies in the weak impurity scattering in our work,
considering the fact that the impurity scattering in the chiral
$p$-wave superconducting state is essentially weak in order to achieve the
superconducting phase.\cite{Sr2,Sr5,IST1,IST2} Consequently, the suppression
of the drive effect from impurity scattering in the previous
work\cite{GOBE1} for $s$-wave case is marginal here.\cite{sscat}

During the optical response with the transverse conical
magnetization, we also find that the triplet vector oscillates as {\small{${\bf
  d}=d(t)\Big(\cos(qx)\sin\theta_{\bf d}(t),-\sin(qx)\theta_{\bf
  d}(t),\cos\theta_{\bf d}(t)\Big)$}}
with the excited space-independent oscillations in the magnitude $d(t)$ 
[Fig.~\ref{figyw7}(a)] and longitude angle $\theta_d(t)$ [Fig.~\ref{figyw7}(b)].
Specifically, as shown in Fig.~\ref{figyw7}(a), the oscillation of the
magnitude (green dotted curve) with frequency $2\Omega$ (referred to
as Higgs mode\cite{Higg1,Higg2,Higg3,Higg4,Higg5,GOBE1}) is dominated by the
drive effect (yellow chain curve), similar to the $s$-wave case.\cite{GOBE1} 

As for the excited triplet-vector rotation (red solid curve) shown in
Fig.~\ref{figyw7}(b), the excited effective magnetic field [seventh term in
Eq.~(\ref{Em2})], shown by blue dashed curve, makes the dominant contribution.
This can be understood from the fact that the triplet vector in the static
situation tends to be parallel to the magnetic
field.\cite{TA1,TA2,TA3,TA4,TA5,TA6,GOBE2} Consequently, in the optical 
response, the excited effective in-plane magnetic field (${\bf h}_{\rm in}^{\rm
  eff}\propto{p_s}{\bf A}_s$), can induce the oscillating in-plane component of
the triplet-vector along ${\bf 
  A}_s=(\cos{qx},-\sin{qx},0)$ with frequency $\Omega$. 
Particularly, this optically excited triplet-vector rotation 
can be enhanced through the enhancement of the effective in-plane magnetic field
by increasing the conical angle, as shown Fig.~\ref{figyw7}(b). 
This prediction provides a scheme to experimentally manipulate the triplet
vector via the optics methods.

\section{SUMMARY AND DISCUSSION}
\label{summary}

In summary, with the impurity scattering included, by extending the
gauge-invariant optical Bloch equation from the   
$s$-wave superconducting state into the chiral $p$-wave state, we 
systematically investigate the AHC induction in the chiral $p$-wave  
superconducting states.
In this study, we focus on THz optical pulse response in GaAs QW in proximity  
to the chiral $p$-wave superconductor Sr$_2$RuO$_4$. However, the
predictions revealed in our study can still shed light on continuous-wave
optical response in the chiral $p$-wave superconductor. Particularly,
in contrast to the well established Kubo formalism,\cite{KB1,KB2,KB3} we provide
a microscopic kinetic description for the AHC induction and reveal a new AHC
induction channel from impurity scattering in our work.

Specifically, we first demonstrate that the intrinsic AHC without magnetic field
is zero as a consequence of the Galilean invariance, in agreement with the 
previous works.\cite{qw,cm,KB1,KB2,KB3} As for the extrinsic AHC, we show that
even at zero temperature, there exists optically excited non-zero normal quasiparticle
current, and a finite anomalous Hall current from
the impurity scattering is obtained, in consistence with the
experiment.\cite{Kerr1} Particularly, the excited normal quasiparticle current
arises from the induced CM momentum of Cooper pairs through the acceleration
driven by the ac electric field, which has been overlooked in the previous
semiclassical approach.\cite{Bo} Specifically, as pointed out in the previous
works,\cite{FF1,FF2,FF3,FF4,FF5,FF6,FF7,FF8,FF9,GOBE1} the induction of a large CM
momentum excites Bogoliubov quasiparticles by breaking the Cooper 
pairs, i.e., forming the blocking region with the markedly suppressed anomalous
correlation, leading to the induced quasiparticle current even at zero temperature.
We find that there exist two AHC induction channels from the
impurity scattering: Channel I, conventional linear channel, in which the AHC is
induced from the skew scattering in the linear response; Channel II, anomalous
nonlinear channel, where the AHC is induced from the Born contribution due to
the broken particle-hole symmetry by the nonlinearly excited quasiparticle
correlation. Both induction channels change sign under 
the opposite chiral characters and vanishes when the superconducting
order parameter approaches zero above the transition temperature, in consistence with
the experiment.\cite{Kerr1} 

Particularly, we point out that Channel I from our
microscopic kinetic description mostly confirms the results of Kubo diagrammatic formalism
of the skew scattering including the electron-density, impurity-density and
order-parameter dependences. Nevertheless, the frequency dependence of AHC at
high frequency in our work ($\sigma^{\rm I}_{xy}\propto{\Omega^{-2}}$) is
different from the Kubo diagrammatic formalism of the skew scattering
($\sigma^{\rm Kubo}_{xy}\propto{\Omega^{-3}}$). This frequency-dependent
difference comes from the different scattering treatments. Specifically, for Kubo
formalism, the skew scattering is treated within perturbation theory by
picking up particular crossed Feynman diagrams.\cite{KB1,KB2,KB3} In our
work, we use the scattering $T$-matrix 
method, which has been widely applied to the kinetic description
in the literature.\cite{Bo,Tm1,Tm2,Tm3,Tm4,Tm5,Tm6} Following the standard
treatment of energy $E$ in the $T$-matrix $T_{\bf kk'}(E)$ in normal
metals ($E=E_F$), we takes $E=E_k$ here, exactly same as the previous work
for superconducting state.\cite{Bo} In the normal state,
these two approaches for the AHC
mostly give inconsistent results with each other and are
considered as complementary techniques,\cite{Tm4} differing
from the longitudinal conductivity in which the Kubo formalism via
picking up the self-consistent ladder Feynman diagrams agrees well with the
kinetic description. For the chiral $p$-wave
superconducting state, the prediction of the AHC from our microscopic kinetic
description agrees qualitatively with the present
experiment,\cite{Kerr1}  similar to the well established Kubo
formalism.\cite{KB1,KB2,KB3} Further 
experiments on the frequency dependence would help to determine which
method better describes the AHC in the chiral $p$-wave superconducting state. 
Nevertheless, to elucidate a complete link between the kinetic description
and Kubo formalism, a gauge-invariant extension for the Kubo formalism should be
developed, since differing from our gauge-invariant description, a special
gauge with zero scalar potential, zero superconducting phase and finite
vector potential is chosen in the Kubo approach in the literature.\cite{KB1,KB2,KB3}
Nevertheless, from the gauge structure revealed by 
Nambu,\cite{gi0} one cannot choose two quantities in the vector potential,
scalar potential, and superconducting phase to be zero.

As for Channel II in our work, to the best of
our knowledge, this exotic channel has long been overlooked in the literature,
due to the difficulty in treating the quasiparticle correlation in the previous
semiclassical approach\cite{Bo} or including the nonlinear effect in Kubo
diagrammatic formalism.\cite{KB1,KB2,KB3} Nevertheless, 
we find that the conventional linear channel (Channel I) only dominates in the
strong impurity interaction, whereas in the weak impurity interaction, the
anomalous nonlinear channel (Channel II) makes the dominant contribution. Based
on this, Channel II may also play an important role
in Sr$_2$RuO$_4$, considering the fact that the impurity scattering in the 
chiral $p$-wave superconducting state is essentially weak in order to achieve
the superconducting phase.\cite{Sr2,Sr5,IST1,IST2}
In addition, we point out that even in the
relatively strong impurity interaction, Channel II can also play an
important role in the pump-probe spectroscopy. Specifically, the pump pulse can
markedly excite the quasiparticle correlation in the nonlinear
response. Consequently, Channel II, which arises from the excited quasiparticle
correlation, will make an important contribution in the probe of Kerr
rotation.  

Furthermore, motivated by the recent observed penetration of the
superconductivity from Sr$_2$RuO$_4$ into the normal metal 
through a ferromagnet in metal/ferromagnet/Sr$_2$RuO$_4$ junctions,\cite{pe4}
we also study the AHC induction in chiral $p$-wave superconducting
QWs in the presence of a magnetization. In our work, we consider a specific
transverse conical magnetization,\cite{TCM1,TCM2,TCM3,TCM4,TCM5} which breaks the  
translational symmetry and hence the Galilean invariance. In this situation, the
intrinsic AHC, attributed from the anomalous velocity, is no longer zero. We
show that this intrinsic AHC 
induction changes sign under the opposite chiral characters and vanishes
when the superconducting order parameter approaches zero. This indicates that the
induced intrinsic AHC is exactly related to the chiral $p$-wave
superconductivity, similar to the extrinsic one from impurity
scattering. Comparison between the intrinsic and extrinsic AHCs from
impurity scattering is also addressed in our work. We show that the intrinsic AHC can
be enhanced though the enhancement of the translational symmetry breaking by
increasing the strength or conical angle of the transverse conical
magnetization. Whereas the extrinsic AHC can be enhanced by increasing impurity
density. 
These dependences provide a scheme to experimentally distinguish the intrinsic
and extrinsic AHC induction channels. 

Finally, the optical response of the chiral $p$-wave superconducting
order parameter is addressed. We reveal that the drive effect of the particles
dominates the optically excited effective chemical potential (related to the
phase of the superconducting order parameter) in the chiral
$p$-wave superconducting state, differing from the $s$-wave case where both the pump
and drive effects play important roles. Moreover, it is also found that   
the optical field excites the rotation of the triplet vector in the presence of
the transverse conical magnetization. This provides a scheme to
experimentally manipulate the triplet vector via the optics methods.

\begin{acknowledgments}
This work was supported by the National Natural Science Foundation of 
China under Grants No.\ 11334014 and No.\ 61411136001.  
\end{acknowledgments}

\begin{appendix}

\section{Gauge invariant optical Bloch equation}
\label{AOBE}

In this section, we construct the optical Bloch equations for chiral $p$-wave
superconducting state in the presence of the
optical field and magnetization. We begin with the generalized Hamiltonian:
\begin{equation}
H_0=[\xi_{-i{\bm \nabla}-e{\bf
    A}(x)\tau_3-{\bf A}^i_s(x){\tilde s}^i\tau_3}+e\phi(x)+{\psi^i_s(x){\tilde s}^i}]\tau_3+{\hat
  \Delta}^{\bf r}_{\bf k}({\bf d}_0), 
\end{equation}
with ${\tilde s}^i={\tilde \sigma}_i/2$ being the spin vector. Here, $\psi^i_s$
denotes the magnetic field and ${\bf A}^i_s$ 
represents the effective SOC coefficient. Under a spin-rotational
transformation $Q(x)=e^{i[{\bm \alpha}(x)\cdot{\bf {\tilde s}}]\tau_3}$, there exists a
non-Abelian gauge\cite{Gr6} structure:
\begin{eqnarray}
\label{F1}
A^i_{s\mu}{\tilde s}^i\tau_3&\rightarrow&{A^i_{s\mu}{\tilde
    s}^i\tau_3-\partial_{\mu}\alpha^i}{\tilde s}^i\tau_3+i[\alpha^i{\tilde
  s}^i,A^k_{s\mu}{\tilde s}^k],~~~~\\
{\bf d}\cdot{\bm \sigma}&\rightarrow&{\bf d}\cdot{\bm \sigma}+i[{\bm \alpha}(x)\cdot{\bf s},{\bf d}\cdot{\bm \sigma}],\label{F2}
\end{eqnarray}
where the four vectors are $A^i_{s\mu}=(\psi^i_s,{\bf A_s}^i)$.\cite{Gr6} It is noted that this
non-Abelian structure in Eqs.~(\ref{F1}) and~(\ref{F2}) is similar to the Abelian structure in
Eqs.~(\ref{gaugestructure1}) and (\ref{gaugestructure2}) for the optical
field. Particularly, the presence of the triplet vector breaks the 
spin-rotational symmetry\cite{Gr6} [Eq.~(\ref{F2})], similar to the fact that the order
parameter breaks the global U$(1)$ symmetry [Eq.~(\ref{gaugestructure2})]. 
Consequently, in the optical Bloch equations, besides the Abelian gauge
invariance, the non-Abelian gauge invariance also has to be retained to
elucidate the complete physical picture, as mentioned in the introduction.
 
Specifically, the optical Bloch equations can be constructed
from the “lesser” Green function $G^{<}_{12}=i\langle\Psi^{\dagger}_2\Psi_1\rangle$,
in which $1/2=x_1/x_2$ and $\langle~\rangle$ denotes the ensemble
average.\cite{Gr1,Gr2,Gr5} With the spin-rotational 
transformation $\Psi(x)\rightarrow{Q(x)}\Psi(x)$, the Green function
$G^<_{12}\rightarrow{Q(x_1)G^<_{12}Q^{\dagger}(x_2)}$. Since in the kinetic equations
in the quasiparticle approximation, only the CM coordinates are
retained,\cite{Gr1}  the 
gauge structure cannot be easily realized in the kinetic equations constructed
from $G^<_{12}$. Nevertheless, as pointed out in Ref.~\onlinecite{GOBE1}, 
the gauge invariance can be retained by introducing
the Wilson line\cite{Gr1,Gr4,Gr5,Gr6,Gr7} to construct the gauge-invariant Green
function, which is constructed as 
\begin{eqnarray}
G^{g<}_{12}&=&{\rm P}e^{-i\int^R_{x_1}dx^{\mu}A_{\mu}\tau_3}e^{-i\int^R_{x_1}dx^{\nu}A^i_{s\nu}{\tilde
  s}^i\tau_3}\nonumber\\
&&\mbox{}{\times}G^<_{12}e^{-i\int^{x_2}_{R}dx^{\nu}A^i_{s\nu}{\tilde s}^i\tau_3}e^{-i\int^{x_2}_{R}dx^{\mu}A_{\mu}\tau_3}. \label{F3}
\end{eqnarray}
In Eq.~(\ref{F3}), $dx^{\mu}=(dt,-d{\bf r})$ and $R=({\bf R},T)=(x_1+x_2)/2$
are the CM coordinates. ``P'' indicates that the line integral is path dependent. By the spin-rotational
transformation $Q(x)=e^{i[{\bm \alpha}(x)\cdot{\bf {\tilde s}}]\tau_3}$, the gauge-invariant Green function is transformed as
$G^{g<}_{12}\rightarrow{Q(R)G^{g<}_{12}Q^{\dagger}(R)}$, in which the transformation operator only
depends on the CM coordinates.

Finally, by choosing the path to be the straight line connecting $x_1$ and
$x_2$,\cite{Gr1,Gr5,GOBE1} via Dyson equations and gradient expansion\cite{Gr1,Gr2,Gr5} (the details of the
derivation can be referred to Ref.~\onlinecite{GOBE1}), the optical Bloch
equations are written as
\begin{eqnarray}
&&\partial_T\rho_{\bf
  k}+i\left[\left(\xi_k+e\phi+\phi_s^i{\tilde s}^i\right)\tau_3+H^{\rm HF}_{\bf k}+{\hat
  \Delta}^{\bf R}_{\bf k}({\bf d_0}),\rho_{\bf
k}\right]\nonumber\\
&&\mbox{}-i\left[\frac{1}{8m}\tau_3,{\bm \nabla}^2_{\bf R}\rho_{\bf
    k}\right]+\left\{\frac{\bf k}{m}\tau_3,{\bm \nabla}_{\bf R}\rho_{\bf
    k}\right\}+\frac{1}{2}\left\{e{\bf E}\tau_3,\partial_{\bf k}\rho_{\bf k}\right\}\nonumber\\
&&\mbox{}-\left[\frac{e{\bm
      \nabla}_{\bf R}\cdot{\bf
    A}}{4m}\tau_3,\tau_3\rho_{\bf k}\right]-\left[\frac{e{\bf
    A}}{2m}\tau_3,\tau_3{\bm \nabla}_{\bf R}\rho_{\bf
    k}\right]\nonumber\\
&&\mbox{}-\left[\frac{\tau_3}{8m},\left[({\bm \nabla}_{\bf R}\cdot{\bf
    A}^i_s){\tilde s}^i\tau_3,\rho_{\bf k}\right]\right]-\left[\frac{\tau_3}{4m},\left[{\bf
    A}^i_s{\tilde s}^i\tau_3,{\bm \nabla}_{\bf R}\rho_{\bf
    k}\right]\right]\nonumber\\
&&\mbox{}+\frac{i}{2}\left[\frac{e{\bf A}}{m}\tau_3,\left\{{\bf
    A}^i_s{\tilde s}^i,\rho_{\bf k}\right\}\right]-\frac{i}{2}\left[\frac{e{\bf k}}{m}\tau_3,\left\{{\bf
    A}^i_s{\tilde s}^i\tau_3,\rho_{\bf k}\right\}\right]\nonumber\\
&&\mbox{}+\frac{1}{2}\left\{i\left[{\bf
    A}^i_s{\tilde s}^i,{
    \phi}^j_s{\tilde s}^j\right]-{\bm \nabla}_{\bf R}{
    \phi}^i_s{\tilde s}^i\tau_3,{\bm \partial_{\bf k}}\rho_{\bf
    k}\right\}+\left[\frac{iA^2}{2m}\tau_3,\rho_{\bf k}\right]\nonumber\\
&&\mbox{}+i\left[\frac{1}{4m}\tau_3,\frac{A^2_s}{4}\rho_{\bf k}+{\bf
    A}^i_s{\tilde s}^i\rho_{\bf k}{\bf
    A}^i_s{\tilde s}^i\right]=\partial_t\rho_{\bf k}|_{\rm scat}.\label{F4}
\end{eqnarray}
Here, ${\bf E}=-{\bm \nabla}_{\bf R}\phi-\partial_t{\bf A}$.
It is noted that with the Abelian [Eqs.~(\ref{gaugestructure1}) and
(\ref{gaugestructure2})] and non-Abelian [Eqs.~(\ref{F1}) and~(\ref{F2})]
gauge structures, Eq.~(\ref{F4}) is gauge invariant for both the Abelian and
non-Abelian gauge transformations.

In our work, in the collinear space, one has $A_{s\mu}{\tilde s}^i=({\bf h_R}\cdot{\bm
  {\tilde \sigma}},0)$, and in the specific gauge with zero superconducting phase, 
Eq.~(\ref{GOBE}) is obtained. In the helical space, with
$A_{s\mu}{\tilde s}^i=\Big(h_0{\tilde \sigma}_z,({\bf A_s}\cdot{\bm {\tilde \sigma}}){\bf e_x}\Big)$, by
neglecting second order of the modulation wave vector, Eq.~(\ref{Em2}) is obtained.

\section{$T$-matrix}
\label{Tmatrix}

We present the scattering $T$-matrix in this section. From the Lippmann-Schwinger
equation:\cite{Bo,Tm1,Tm2,Tm3,Tm4,Tm5,Tm6} 
\begin{equation}
T_{\bf kk'}(E)=z_iV_{\bf k-k'}\tau_3+\sum_{\bf k_1}z_iV_{\bf k-k_1}\tau_3G_{\bf
    k_1}(E)T_{\bf k_1k'},
\end{equation}
the scattering $T$-matrix is given by
\begin{eqnarray}
\label{Tsum}
&&T_{\bf kk'}(E)=z_iV_{\bf k-k'}\tau_3+\sum_{\bf k_1}z_iV_{\bf k-k_1}\tau_3G_{\bf
    k_1}(E)z_iV_{\bf k_1-k'}\tau_3\nonumber\\
&&\mbox{}+\sum_{\bf k_1k_2}z_iV_{\bf k-k_1}\tau_3G_{\bf
    k_1}(E)z_iV_{\bf k_1-k_2}\tau_3G_{\bf
    k_2}(E)z_iV_{\bf k_2-k'}\tau_3\nonumber\\
&&\mbox{}+...,
\end{eqnarray}
with Green function {\small{$G_{\bf
    k_1}(E)=\frac{1}{E-H_{0{\bf
      k}}}=\frac{E-\xi_k\tau_3-\Delta^{\bf R}_{\bf k}({\bf d}_0)}{E^2-E_k^2}$}}.

It is noted that the impurity interaction $z_iV_{\bf kk'}$ has the weak
momentum dependence, since the screening constant $\kappa$ in our study is
compared to the Fermi vector. Consequently, Eq.~(\ref{Tsum}) approximately
becomes: 
\begin{eqnarray}
\label{Tsumf}
&&T_{\bf kk'}(E){\approx}z_iV_{\bf k-k'}\tau_3+z_iV_{\bf k-k'}\tau_3\left[\sum_{\bf k_1}G_{\bf
    k_1}(E)\right]z_iV_{0}\tau_3\nonumber\\
&&\mbox{}+z_iV_{\bf k-k'}\tau_3\left[\sum_{\bf k_1}G_{\bf
    k_1}(E)\right]z_iV_{0}\tau_3\left[\sum_{\bf k_2}G_{\bf
    k_2}(E)\right]z_iV_{\bf 0}\tau_3\nonumber\\ 
&&\mbox{}+...=\frac{z_iV_{\bf
  k-k'}\tau_3}{1+i\tau_3\pi\nu{z_iV_0}E/\sqrt{E^2-\Delta^2_0}}, 
\end{eqnarray}
similar to the previous work for the short-range impurity interaction.\cite{Bo}
For quasiparticles with energy $E=E_k>\Delta_0$, the scattering $T$-matrix used in Sec.~\ref{model} is obtained. 

\section{Normal quasiparticle current}
\label{lrr}

\begin{figure}[htb]
  {\includegraphics[width=7.0cm]{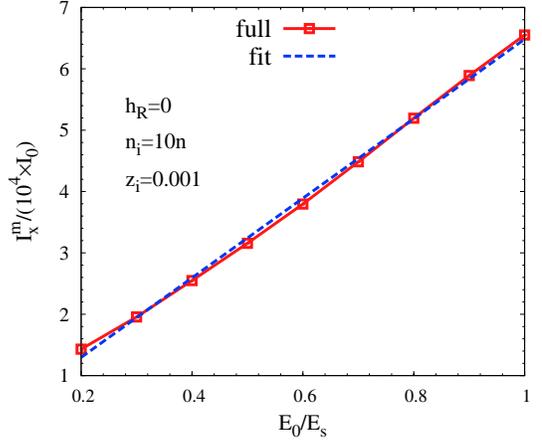}}
\caption{(Color online) Maximum of the excited normal quasiparticle current versus
  electric-field strength. Blue dashed curve: fitted
  result by using $I_{x}\propto{E_0}$. } 
  
\label{figyw8}
\end{figure}

In this part, we show that the optically excited normal quasiparticle current
always lies in the linear-response (i.e., small-$E_0$) regime in our
investigation.  This can be clearly seen from Fig.~\ref{figyw8}, where it is
shown that the maximum of the excited normal
quasiparticle current increases linearly with the electric-field strength. 

\section{Derivation of Eq.~(\ref{TCMin})}
\label{Din}

We derive Eq.~(\ref{TCMin}) in this part. By neglecting the effective chemical
potential, HF and impurity scattering terms, in the linear response,
we first transform Eq.~(\ref{Em2}) from the particle space into the quasiparticle one as: 
\begin{eqnarray}
&&i\Omega\rho^{\tilde q}_{\bf
    k}+i[(E_k+h_0{\tilde \sigma}_z)\tau_3,\rho^{\tilde q}_{\bf
    k}]+\frac{1}{2}\left\{eE_0t_3,[U^{\dagger}_{\bf
      k}\partial_{\bf k}U_{\bf k},\rho^{\tilde q}_{\bf 
    k}]\right\}\nonumber\\
&&\mbox{}+\frac{1}{2}\left\{{({\bf c_s}\cdot{\bm {\tilde
        \sigma}})}t_3,\partial_{k_x}\rho^{\tilde q}_{\bf
    k}+[U^{\dagger}_{\bf
      k}\partial_{\bf k}U_{\bf k},\rho^{\tilde q}_{\bf 
    k}]\right\}\nonumber\\
&&\mbox{}-\left\{\frac{ik_x}{2m}t_3,\left[({\bf
      A_s}\cdot{\bm {\tilde \sigma}})t_3,\rho^{\tilde q}_{\bf k}\right]\right\}=0.
\label{DT1}
\end{eqnarray}
Here, {\small{$({\bf c_s}\cdot{\bm {\tilde \sigma}})\tau_3=[({\bf A_s}\cdot{\bm
  {\tilde \sigma}}),h_0{\tilde \sigma}]$}} and {\small{$\rho^{\tilde q}_{\bf k}=U_{\bf k}\rho^h_{\bf k}U_{\bf
  k}$}}. In Eq.~(\ref{DT1}), we have neglected the in-plane triplet vector and the
excited effective in-plane magnetic field [seventh term in
Eq.~(\ref{Em2})] due to the large out-of-plane
triplet vector and small $p_s$, respectively.

Since the conical angle $\sin(\theta/2)$ is a small quantity in our work, one can
expand the density matrix as $\rho^{\tilde q}_{\bf k}=\sum_l\rho^{l}_{\bf k}$
with $l$ denoting the order of  
conical angle, and Eq.~(\ref{DT1}) becomes: 
\begin{eqnarray}
&&\Omega\rho^{l}_{\bf
    k}+[(E_k+h_0{\tilde \sigma}_z)\tau_3,\rho^{l}_{\bf k}]-\frac{k_x}{2m}\left\{t_3,\left[({\bf
      A_s}\cdot{\bm {\tilde \sigma}})t_3,\rho^{l-1}_{\bf k}\right]\right\}\nonumber\\
&&\mbox{}-\frac{i}{2}\left\{({\bf c_s}\cdot{\bm {\tilde \sigma}})t_3,\partial_{k_x}\rho^{l-1}_{\bf
    k}+[{B}_{\bf k},\rho^{l-1}_{\bf 
    k}]\right\}=0.
\label{DT2}
\end{eqnarray}
From above equation, one has $\rho^{l=0}_{\bf k}=-iD_1k_x\tau_0$, $\rho^{l=1}_{\bf k}=D_1({\bf c}_s\cdot{\bm
    {\tilde \sigma}})t_{\Omega}$ and $\rho^{l=2}_{\bf k}=D_2k_x{\tilde
    \sigma}_z\tau_3t_{\Omega}$ in the lowest three orders. Here,
  $D_1=eE_0\frac{2u^2_kv^2_k}{{\Omega}mE_k}$, $D_2=\frac{2iD_1A_sc_s}{m\Omega}$ and
  $t_{\Omega}=\frac{u^2_k-v^2_k}{\Omega}\tau_3+\sum_{\pm}\frac{2u_kv_k}{\Omega{\pm}2E_k}\sigma_1\tau_{\pm}$. Then,
  $\rho^{l=3}_{\bf k}$ is obtained as
\begin{eqnarray}
\rho^{l=3}_{\bf k}&=&D_3^0\left(\begin{array}{cc}
\frac{2u_kv_k({{\bf e_z}\times{\bf c_s}})\cdot{\bm \sigma}}{\Omega^2-4E^2_k} &
\frac{(v_k^2-u_k^2)({{\bf e_z}\times{\bf c_s}})\cdot{\bm \sigma}\sigma_1}{(\Omega+2E_k)^2}\\
\frac{(v_k^2-u_k^2)\sigma_1({{\bf e_z}\times{\bf c_s}})\cdot{\bm \sigma}}{(\Omega-2E_k)^2}& -\frac{2u_kv_k({{\bf e_z}\times{\bf c_s}})\cdot{\bm \sigma}^*}{\Omega^2-4E^2_k}\end{array}\right)\nonumber\\
&&-\frac{iD^0_3k_xk_y}{k^2}\left(\begin{array}{cc}
\frac{D^{1}_3({{\bf e_z}\times{\bf c_s}})\cdot{\bm \sigma}}{\Omega} &
\frac{D^{+}_3({{\bf e_z}\times{\bf c_s}})\cdot{\bm \sigma}\sigma_1}{\Omega+2E_k}\\
-\frac{D^{-}_3\sigma_1({{\bf e_z}\times{\bf c_s}})\cdot{\bm \sigma}}{\Omega-2E_k}& -\frac{D^{1}_3({{\bf e_z}\times{\bf c_s}})\cdot{\bm \sigma}^*}{\Omega}\end{array}\right)\nonumber\\
&&+2ih_0D_3^0\left(\begin{array}{cc}
\frac{2u_kv_k({\bf c_s}\cdot{\bm \sigma})}{\Omega(\Omega^2-4E^2_k)} &
\frac{(v_k^2-u_k^2)({\bf c_s}\cdot{\bm \sigma})\sigma_1}{(\Omega+2E_k)^3}\\
\frac{(v_k^2-u_k^2)\sigma_1({\bf c_s}\cdot{\bm
  \sigma})}{(\Omega-2E_k)^3}& -\frac{2u_kv_k({\bf c_s}\cdot{\bm
  \sigma}^*)}{\Omega(\Omega^2-4E^2_k)}\end{array}\right)\nonumber\\
&&+\frac{2h_0D^0_3k_xk_y}{k^2}\left(\begin{array}{cc}
\frac{D^{1}_3({\bf c_s}\cdot{\bm \sigma})}{\Omega^2} &
\frac{D^{+}_3({\bf c_s}\cdot{\bm \sigma})\sigma_1}{(\Omega+2E_k)^2}\\
-\frac{D^{-}_3\sigma_1({\bf c_s}\cdot{\bm
  \sigma})}{(\Omega-2E_k)^2}& -\frac{D^{1}_3({\bf c_s}\cdot{\bm
  \sigma}^*)}{\Omega^2}\end{array}\right),~~~~~~
\label{DT4}
\end{eqnarray}
with $D_3^0=\frac{4u_kv_kD_2E_k}{\Omega}$,
$D^1_3=\frac{4u_kv_kE_k(u_k^2-v_k^2)}{\Omega^2-4E^2_k}$ and
$D^{\pm}_3=\frac{4u_k^2v_k^2\Omega}{\Omega^2-4E^2_k}+\frac{(u_k^2-v_k^2)^2}{\Omega\pm2E_k}$.

Finally, from Eqs.~(\ref{DT2}) and ~(\ref{DT4}), the induced anomalous Hall
current reads:
\begin{eqnarray}
I_y&=&\frac{e}{2m}\int{d{\bf k}}{\rm Tr}(k_y\rho^{\tilde q}_{\bf k})\approx\frac{e}{2m}\int{d{\bf k}}{\rm Tr}(k_y\rho^4_{\bf k})\nonumber\\
&=&\frac{e}{\pi}\int{d\varepsilon_k}D_4\left[\frac{2\xi_k}{E_k}\frac{D^1_3}{\Omega^2}+\sum_{\pm}\frac{2u_kv_kD^{\pm}_3}{(\Omega\pm2E_k)^2}\right], ~~~~
\label{DT5}
\end{eqnarray}
which becomes Eq.~(\ref{TCMin}) at high frequency.

\end{appendix}


\begin{thebibliography}{0}
\bibitem{ep} The experimentally observed Kerr effect
was also considered within clean multiband models.\cite{mu1,mu2,mu3,mu4,mu5}
Nevertheless, Kallin pointed out\cite{Sr5} that the multiband model deviates
from the present approximation and discussion where it is assumed that the
superconductivity in Sr$_2$RuO$_4$ is predominantly on the $\gamma$ band with
weak induced superconductivity on the $\alpha$ and $\beta$ bands.
\bibitem{sscat} For $s$-wave case, we also find that the drive effect dominates
  the optically excited effective chemical potential in the weak impurity
  interaction (not shown), which is absent in the previous work.\cite{GOBE1} This is in marked
  contrast to the conclusions from the Liouville equation\cite{sw1,sw2,sw3} or Bloch
  equation\cite{sw4,sw5,sw6,sw7,sw8,sw9} in the literature, in which the drive effect is
  overlooked.
\end{thebibliography}
\end{document}